\documentclass[pra,superscriptaddress,onecolumn,amsmath,amssymb]{revtex4}
\usepackage{amsfonts}
\usepackage{amssymb}
\usepackage{mathtools}
\usepackage{graphicx}
\usepackage[english]{babel}
\usepackage{subfigure}
\usepackage{mathrsfs}
\usepackage{color}
\usepackage[normalem]{ulem}
\usepackage{natbib}
\usepackage{bbold}
\usepackage{placeins}
\usepackage{braket}
\usepackage{soul,xcolor}
\usepackage{epstopdf}
\usepackage{tabu}
\usepackage{array}
\usepackage{bm}
\usepackage{upgreek}
\usepackage{multirow}
\usepackage[utf8]{inputenc}
\usepackage[colorlinks = truelinkcolor = blue, urlcolor  = blue, citecolor = blue, anchorcolor = blue]{hyperref}
\hypersetup{
	colorlinks   = true, 
	urlcolor     = black, 
	linkcolor    = blue, 
	citecolor   = blue 
}

\makeatletter
\begin{document}
	
	\title{Effect of memory on the violation of Leggett-Garg inequality}

\author{Javid Naikoo}
\email{naikoo.1@iitj.ac.in}
\affiliation{Indian Institute of Technology Jodhpur, Jodhpur 342011, India}

\author{Subhashish Banerjee}
\email{subhashish@iitj.ac.in}
\affiliation{Indian Institute of Technology Jodhpur, Jodhpur 342011, India}

\author{ R. Srikanth}
\email{srik@poornaprajna.org}
\affiliation{Poornaprajna Institute of Scientific Research, Bangalore - 562164, India}

\begin{abstract}
		\noindent 
The Leggett-Garg inequalities  impose restrictions on the values taken by some combinations of the two-time correlation functions of observables in order to be explainable by a noninvasive-realist classical model. While in the unitary dynamics, it is straightforward to compute these correlation functions, open system effects bring in  subtleties. Specifically, for  non-Markovian dynamics, which   involves setting up of system-bath correlations, the Leggett-Garg measurements disrupt these correlations, making a full system-bath Hamiltonian approach natural. However, here we point out that the problem can also be dealt with from a reduced dynamics perspective. The key point is that the noise superoperator acting on the system must be suitably updated after measurement interventions. Also considered is the effect of Markovian versus non-Markovian behavior as well as classically non-Markovian processes on the violation of  Leggett-Garg inequalities.
\end{abstract}
	\maketitle

\section{Introduction}
The study of nonclassical correlations has not only turned out to be an important tool in probing the basic features which make a quantum system different from a classical system, but has also provided potential resources for future quantum technologies. The nonclassical correlations can be quantified in many ways. The celebrated Bell inequalities \cite{bell1964einstein} serve as a test for the local realism. Quantum  steering \cite{PhysRev.47.777} allows one party to change the state of the other by local measurements. The nonseparability of the state of a system is probed by so called entanglement witnesses \cite{Schordinger1}. Quantum discord is another measure of nonclassical correlation which can even exist in systems which are not entangled \cite{discordOliverZurek,DiscordSB}. These spatial quantum correlations have been a subject matter of many theoretical \cite{banerjee2016quantum,alok2016quantum,banerjee2015quantum,Alok:2015iua,indranilsb,SbRaviSrik,SbRaviSrik2,Arend,Kaer,Imran2015,Imran2016,WeiJiang2018,naikoo2018probing} and experimental works \cite{aspect1981experimental,tittel1998experimental,lanyon2013experimental,weihs1998violation}.

Temporal quantum correlations, which exist between different measurements made on a single system at different times, have attracted lot of attention in recent years. Prominent among these are the Leggett-Garg inequalities (LGIs). 
In their different forms, LGIs   has been analyzed in various  theoretical  \cite{barbieri2009multiple,avis2010leggett,lambert2010distinguishing,lambert2011macrorealism,emary2013leggett,kofler2013condition,leggett1985quantum,montina2012dynamics,Emary2012,EmaryDecoh,Naikoo:2017fos,javidLGImeson,mal2016quantum,Naikoo2018} and experimental works 
\cite{palacios2010experimental,goggin2011violation,xu2011experimental,dressel2011experimental,suzuki2012violation,athalye2011investigation,katiyar2013}.  

The  two important macroscopic notions on which LGIs are formulated are \textit{macrorealism} and \textit{non-invasive measurability} \cite{leggett1985quantum,emary2013leggett}. Macrorealism means that a system, which has available to it two or more macroscopically distinct states, must be in one of these states at any given time. Non-invasive measurability means that the act of measurement reveals the state of the system without disturbing its future dynamics. Both these assumptions are not respected by quantum systems; the superposition principle violates the first and the collapse of the wavefunction under measurement defies the second. With these two assumptions, the simplest form of LGI in terms of  the LG parameter $K_3$
\begin{equation}\label{Eq-K3}
K_3 = C(t_0, t_1) + C(t_1, t_2) - C(t_0, t_2),
\end{equation}
is given by $-3\le K_3 \le 1$. Here $C(t_i, t_j) = \langle Q(t_i) Q(t_j)\rangle$ is the two-time correlation for the dichotomic observable $Q(t)=\pm1$. 
The two-time correlations $C_{ij}$ appearing in Eq. (\ref{Eq-K3}) can be written in terms of the conditional probabilities as 
\begin{align}\label{Cij}
C(t_i, t_j) &= p(^{+}t_i)q(^{+}t_j|^{+}t_i) - p(^{+}t_i)q(^{-}t_j|^{+}t_i)  - p(^{-}t_i)q(^{+}t_j|^{-}t_i) + p(^{-}t_i)q(^{-}t_j|^{-}t_i),
\end{align}
where $p(^{a}t_i)$ is the probability of obtaining the  result $a = \pm1$ at $t_i$, and $q(^{b}t_j|^{a}t_i)$ is the conditional probability of getting result $b=\pm 1$ at time $t_j$, given that result $a = \pm 1$ was obtained at $t_i$. 

Suppose   Alice  and  Bob  measure   observables  $\hat{A}$  and
$\hat{B}$, obtaining outcomes $a, b \in \{\pm1\}$. Then, for any input
state $\rho$, one finds 
\begin{align}
P_{ab|\hat{A}(t_i)\hat{B}(t_j)}  &=   \textrm{Tr}\left(\frac{1  +  b\hat{B}(t_j)}{2}
\frac{1 + a\hat{A}(t_i)}{2}\rho \frac{1 + a\hat{A}(t_i)}{2}\right) 
\nonumber\\ 
&= \frac{1}{4}
+  \frac{a}{4}\textrm{Tr}(\hat{A}(t_i)\rho)                +
\frac{b}{8}\textrm{Tr}(\hat{B}(t_j)\rho)                                 
+
\frac{ab}{8}\textrm{Tr}(\{\hat{A}(t_i),\hat{B}(t_j)\}\rho)                      +
\frac{b}{8}\textrm{Tr}(\hat{A(t_i)}\hat{B}(t_j)\hat{A}(t_i)\rho),
\label{eq:tjempo}
\end{align}
from  which it  follows  that the  correlator  $C(t_i, t_j) \equiv  \langle
\hat{A}(t_i)\hat{B}(t_j)   \rangle   =  \sum_{a,b}   abP_{ab|\hat{A}(t_i)\hat{B}(t_j)}   =
\frac{1}{2}\langle \{\hat{A}(t_i)\hat{B}(t_j)\}  \rangle = \vec{A}(t_i)\cdot\vec{B}(t_j)$,
where   $\hat{A}(t)   =    \vec{A}(t)\cdot\vec{\sigma}$   and   $\hat{B}(t)   =
\vec{B}(t)\cdot\vec{\sigma}$  \cite{aravinda2012general,fritz}, where $\sigma$ are the Pauli matrices.  Thus, the
correlators $C_{ij}$  are independent of  the input state, if  the two
measurements are projective. In the context of LGIs, $\hat{A}$ and $\hat{B}$ would be the observable $\hat{Q}$ at different times $t_i$ and $t_j$; similar conclusions follow.

It emerges from our work that the intervening
noise between  two measurements is  relevant for the evolution of the LG parameter.  This can  be understood
equally well by  absorbing the noise into the  measurements, which can then be regarded  as  a  noise-induced POVM \cite{kumariprob},  and  no  longer  projective
measurements.

In this work, we will study the effect of non-Markovianity on temporal correlations, in particular, as part of a test for LGI. From a quantum information theoretic perspective, non-Markovianity has of late been studied by the (not always equivalent) criteria of (CP) divisibility and distinguishability \cite{rivas2014quantum}. In particular, non-Markovianity according to the former criterion manifests as the fact that the \textit{intermediate map} (i.e., the dynamical map that propagates an intermediate earlier state to a later state) acting on the density operator is not-completely-positive (NCP) \cite{pradeep2}. As a result, the intermediate time evolution of the density operator is no longer given by the Kraus operator-sum representation. Instead, the operator sum-difference representation must be employed \cite{omkar2015operator}, wherein the trace-preserving NCP map is represented as the difference of two CP maps. 

In the context of temporal  correlations, that they will also be affected by non-Markovianity, as are general quantum phenomena, is not surprising.  Indeed, a  sufficient
but not necessary  measure for non-Markovianity in  terms of a
temporal      steerable      weight      is      given      in
\cite{chen2016quantifying}.  However, there is an important fact to be recognized here, which is that non-Markovianity in general involves setting up system-bath correlations, even through the system and bath may be initially uncorrelated. Therefore, the intervention of measurement that is done to produce temporal correlations, will in general re-prepare the environment also, just as it re-prepares the system. Hence, correlations based on a subsequent measurement will be subject, in general, to a different noisy channel than the first measurement, and furthermore, would depend on the output of the preceding measurement. The above observation seems to be implicitly present in existing treatments of temporal correlations under scenarios where the assumption that the system and environment retain a factorized form has to be given up in some way. In such works, typically the joint system-environment evolution is considered, rather than the reduced dynamics, in order to derive the system correlation functions.

In recent times, there have been a number of works that compute the two-time correlation functions for non-Markovian dynamics. For example, the evolution equations for the two-time correlation functions for non-Markovain evolution in the case of weak system-environment coupling was studied in \cite{goan2011non}, employing the full system-environment Hamiltonian.    In particular, with regard to the question of the LGI violation in the context of non-Markovian noise,  building on \cite{goan2011non},  the LGI violations for a two-level system under non-Markovian dephasing was studied in  \cite{chen2014investigating}. A similar problem for the Jaynes-Cummings model was discussed in  \cite{BAN20172313}. The common theme in these works is to start from the full unitary evolution and then derive the evolution equations for the correlation functions  using the appropriate limits. In contrast to these works, here will explore the direct use of the system dynamics for studying LGI violation, indicating the scope and constraints of this approach. In a related vein: the failure of the quantum regression hypothesis (QRH) \cite{swain}, which deals with multi-time correlation functions, also captures a traditional idea of quantum non-Markovianity \cite{guarnieri2014quantum}.

As noted above, because under non-Markovianity, system measurements can disturb the bath, and hence care must be exercised in computing two-time correlations if the reduced dynamics alone is used. Here, we study LGI violation in the non-Markovian regime which, to our knowledge, is the first instance where this is done using the system's reduced dynamics.  We argue that a purely reduced dynamics approach can be adopted, with the proviso that the noise is suitably updated in an outcome-dependent manner after the first (and subsequent) intervention(s). 

The plan of  this work is as follows:   Section (\ref{sec:JCmodel}) is  devoted to  a  description of a simple non-Markovian model and its characterization. Further, in order to ascertain the impact of Markovian versus non-Markovian behavior as well as to understand quantum and classical non-Markovian effects on the LGI, we consider two models, namely, the \textit{phase damping} (PD) \cite{NC} and the quantum semi-Markov processes \cite{budini}, respectively. The corresponding LGIs, in the context of these models, are discussed in Sec. (\ref{LGIs}).   Conclusion of the work is presented    in  Sec. (\ref{conclusion}).

\section{Noise Models}\label{sec:JCmodel}

Here we consider a few noisy models with the subsequent aim of studying the LGI violation.

\subsection{A Simple Model}
Given times $t_2 > t_1 > t_0$ during the evolution of an open system, suppose a projective measurement is performed at time $t_1$.  If the environment is (approximately) stationary during the interval $[t_0,t_2]$, then the same channel can be considered as acting in the intervals  $(t_1, t_2) $ and $(t_0, t_1)$.   Let the Hilbert spaces of the system and environment be denoted by $\mathcal{H}_S$ and $\mathcal{H}_E$, respectively; with initial states $\ket{\psi_S} \in \mathcal{H}_S$ and $\ket{\psi_E} \in \mathcal{H}_E$, respectively \cite{breuer}. The combined state $\ket{\psi_S} \otimes \ket{\psi_E}$ lives in the tensor product space $\mathcal{H}_S \otimes \mathcal{H}_E$. The total dynamics is given by unitary $(U)$, and  would in general entangle the system and environment degrees of freedom such that the reduced dynamics, say from $t_0$ to $t_1$, is described by the Kraus operators $K_\mu (t_1 - t_0) = \langle e_\mu | U(t_1 - t_0) | \psi_E \rangle$, where $\{\ket{e_\mu}\}$ is a basis for the environment \cite{kraus1983states}. An act of measurement at time $t_1$ would collapse the system in an eigenstate of the projector and simultaneously modify the state of  environment to $\ket{\psi_E^\prime}$. The new Kraus operators, governing the dynamics from $t_1$ to $t_2$ would be $K^\prime_\mu (t_2 - t_1) = \langle f_\mu | U(t_2 - t_1) | \psi_E^\prime \rangle$, where $\ket{f_\mu} = e^{i \xi} \ket{e_\mu}$ is a new environment basis. Assuming that the environment state changes only by a global phase $e^{i \chi}$, i.e., $\ket{\psi_E^\prime} = e^{i \chi} \ket{\psi_E}$,  we have $K_\mu^\prime (t_2 - t_1) = e^{i (\xi - \chi)} K_\mu (t_1-t_0)$. Thus the two Kraus operators differ only by a global phase factor and hence describe the same dynamics, i.e., the reduced dynamics they produce has the same time dependence.

Here, a crucial assumption made was that the act of measurement changes the state of environment at most by a global phase. We will now illustrate, using a simple model, that such an assumption does not hold for non-Markovian dynamics and one needs to update the post-measurement map depending upon the measurement outcome. 	 If the system-bath interaction is a product of local unitaries (i.e., it is not an entangling operation), then the system dynamics is necessarily Markovian and CP-divisible. Therefore, if the system dynamics is CP-indivisible, then in general the system-bath interaction is an entangling operation, that would generate entanglement between the system and bath. In the former case, this allows for a clean separation between the system and reservoir time scales, which is not so in the later case where the system-bath interaction is an entangling operation. Clearly, the above argument  will no longer hold, requiring the system dynamics to be modified post-measurement. To see how one must modify it, we consider a simple model of system-bath interaction.
	
    We consider a  model which is a two qubit system, where one qubit is the environment $E$ and the other is the system $S$, such that the entanglement between the two qubits shows up as noise in the reduced dynamics of the first qubit. For capturing the main conceptual points, we  choose a simple (but non-trivial) environment to highlight the point that the environment itself is reset after a measurement in the non-Markovian situation.  Let us denote the initial state of the system and environment  by 
	\begin{equation}\label{eq:states}
	\ket{\psi_S} = (\ket{0_S} + \ket{1_S})/\sqrt{2} \qquad {\rm and}\qquad \ket{\psi_E} = (\ket{0_E} + \ket{1_E})/\sqrt{2},
	\end{equation}
	where the subscripts $S$ and $E$ correspond to system and environment, respectively. Let us assume a separable  state at time $t=0$, that is,  $\ket{\psi(0)} = \ket{\psi_S} \otimes \ket{\psi_E}$. We adopt the Hamiltonian  (with $\hbar = 1$)
	\begin{equation}\label{eq:H}
	H  = \omega \Big(| 01 \rangle \langle 10 | +  | 10 \rangle \langle 01 | \Big),
	\end{equation}
 which is reminiscent of the Jaynes-Cummings Hamiltonian, where the optical mode, restricted to the single excitation subspace, is treated like a two-level system (qubit). Then, $\omega$ can be treated as the frequency of the Rabi-like oscillations which happen between the two Bell states $\frac{1}{\sqrt{2}}(\ket{01} \pm \ket{10})$. 
As a consequence, the time evolution generated by  unitary operator  $	U(t)  = e^{-i H t}$ corresponds to an entangling operation between the two qubits.
	 Let us define the   density matrices corresponding to system $\rho_{S}  = \ket{\psi_S}\bra{\psi_S}$, environment   $\rho_{E}  = \ket{\psi_E}\bra{\psi_E}$, and the composite state   $\rho_{SE} (t) = \ket{\psi(t)} \bra{\psi(t)}$.\bigskip

\begin{figure}
 	\centering
 	\includegraphics[width=80mm]{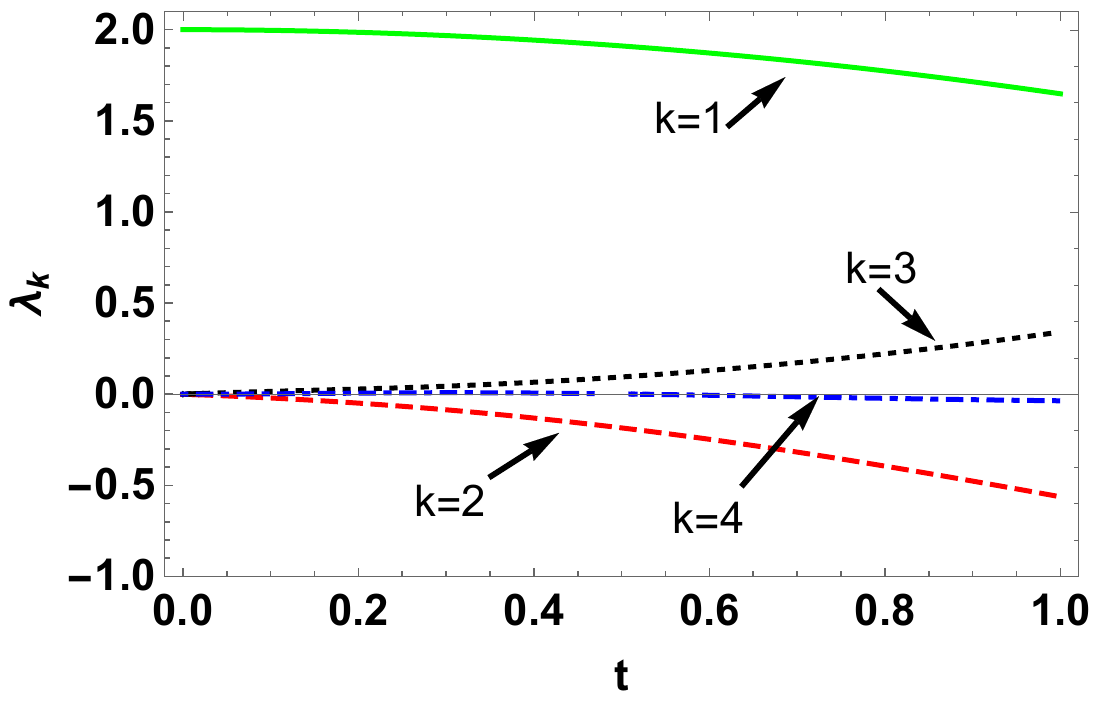}
 	\caption{(color online) Eigenvalues $\lambda_k$ ($k=1,2,3,4$) of the Choi matrix $\bm{B}_{pr; qs} (t, t/2)$. Negative eigenvalues indicate that the map is NCP.}
 	\label{fig:EVs}
 \end{figure}

	 	\textit{Characterization of non-Markovian dynamics:} 
	Here, we investigate the non-Markovian features of the above mentioned model  by studying Sudarshan's $\bm{A}$ and $\bm{B}$ dynamical maps \cite{Sudarshan}. 
	 The dynamics of an open system involves mapping an initial input state $\bm{\rho}(0)$ to an output state $\bm{\rho}(t)$ at a given time $t$ by a linear map $\bm{A}(t,0)$. This  is done by vectorizing the reduced system density matrix $\bm{\rho}_S$, obtained by tracing over the environment $E$, such that $\bm{\rho}_S^\prime = \bm{A}.\bm{\rho}_S$, or $\bm{\rho}_{p,q}(t_0) \rightarrow \bm{\rho}^{\prime}_{p,q}(t) = \bm{A}_{pq;rs}(t, t_0) \bm{\rho}_{r,s}(t_0)$.
	 
	 \begin{equation}\label{eq:Bmat}
	 \bm{A}(t,0) =\left(
	 \begin{array}{cccc}
	 	\frac{1}{4} (\cos (2 t \omega )+3) & \frac{1}{4} i \sin (2 t \omega ) & -\frac{1}{2} i \sin (t \omega ) & \cos (t \omega ) \\
	 	\frac{1}{2} i \sin (t \omega ) & \cos (t \omega ) & 0 & -\frac{1}{2} i \sin (t \omega ) \\
	 	-\frac{1}{2} i \sin (t \omega ) & 0 & \cos (t \omega ) & \frac{1}{2} i \sin (t \omega ) \\
	 	\frac{1}{2} \sin ^2(t \omega ) & -\frac{1}{4} i \sin (2 t \omega ) & \frac{1}{4} i \sin (2 t \omega ) & \frac{1}{4} (\cos (2 t \omega )+3) \\
	 \end{array}
	 \right). 
	 \end{equation}
 In order to show the CP indivisibility of the map, we divide the time evolution between $(0,t)$ into interval $(0, t/2)$ and $(t/2,t)$, such that $\bm{A}(t, t/2) = \bm{A}(t,0) \bm{A}^{-1}(t/2,0) $.   One can then construct the $\bm{B}(t, t/2)$ map, which is basically the Choi matrix, by using 
 \begin{equation}\label{eq:Bmap}
 \bm{B}_{pr; qs} (t, t/2) = \bm{A}_{pq; rs} (t, t/2).
 \end{equation}
 The eigenvalues of this matrix  are plotted in Fig. (\ref{fig:EVs}). The negative eigenvalues indicate CP-indivisibility of the map.
  
  	In fact, it is possible to show that the map is P-indivisible. For that it is enough to show that the evolution under this map leads to increase in the distinguishability of two states. This can be shown by looking at the behavior of  trace distance function between two orthogonal states subjected to the map $\Phi^o$ described by Eq. (\ref{eq:Phi}) below.   Consider two orthogonal states $\rho_0(0) = \ket{0}\bra{0}$ and $\rho_1(0) = \ket{1}\bra{1}$, evolved under this map to $\rho_0(t) = \Phi^o[\rho_0(0)] $ and  $\rho_1 (t) = \Phi^o[\rho_1(0)]$, respectively.
	 
	 The trace distance between these states  is defined as $ {\rm TD} = \frac{1}{2} \sum_k |\eta_k|$, where $\eta_k$ are the eigenvalues of matrix $ \rho_0(t)  - \rho_1(t)$. We have
	 \begin{equation}\label{eq:TD}
	 {\rm TD} = \frac{1}{2}\sqrt{\frac{7 + \cos(\omega t)}{2}}. 
	 \end{equation}
	 It is clear that TD is an oscillating function of time. The recurrent behavior of TD is a signature of P-indivisibility of the map, and could be interpretted as the backflow of information updating the system dynamics.\bigskip

	\begin{figure}
		\centering
		\includegraphics[width=80mm]{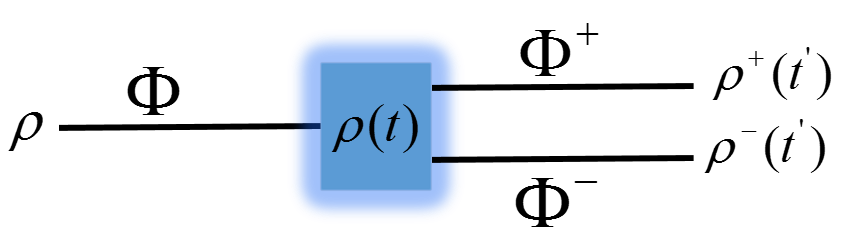}
		\caption{(color online) A measurement of a dichotomic observable on $\rho(t)$ would be followed by two possible dynamics depending on its outcome. The map $\Phi^o$ (Eq. (\ref{eq:Phi}))  would be replaced by  $\Phi^{+}$ and $\Phi^{-}$ (Eq. (\ref{eq:Phipm})) depending on whether the outcome is $+1$ or $-1$, respectively. }
		\label{fig:diag}
	\end{figure}
	   
	  \textit{Reduced dynamics:}  The reduced state of the system can be obtained by tracing over the environment. Denoting the set of  basis states of the environment as $\{\ket{e_\mu}$\}, we have $	\rho_S(t)  = \sum_{\mu} \mathcal{K}_\mu \rho_{S} \mathcal{K}_\mu^\dagger$, where $ \mathcal{K}_{\mu} = \langle e_\mu | U (t) | \psi_E \rangle$ are the Kraus operators.      With the Hamiltonian given by Eq. (\ref{eq:H})   and the environment state given in Eq. (\ref{eq:states}) (the environment basis states $\{\ket{e_\mu}= \ket{0_E}, \ket{1_E}\}$), we obtain
	\begin{align}\label{eq:K}
	\mathcal{K}_0 = \left(
	\begin{array}{cc}
	\frac{1}{\sqrt{2}} & 0 \\
	-\frac{i \sin (\omega t)}{\sqrt{2}} & \frac{\cos (\omega t)}{\sqrt{2}} \\
	\end{array}
	\right),\qquad {\rm and}\qquad \mathcal{K}_1  = \left(
	\begin{array}{cc}
	\frac{\cos (\omega t)}{\sqrt{2}} & -\frac{i \sin ( \omega t )}{\sqrt{2}} \\
	0 & \frac{1}{\sqrt{2}} \\
	\end{array}
	\right),
	\end{align}
	satisfying the completeness relation $\mathcal{K}_0^\dagger  \mathcal{K}_0 + \mathcal{K}_1^\dagger  \mathcal{K}_1 = \mathbb{1}$.
	\begin{equation}\label{eq:Phi}
	\Phi^o[\rho_S(0)] =\sum\limits_{\mu} \mathcal{K}_{\mu}  \rho_S(0) \mathcal{K}_\mu^\dagger.
	\end{equation}
It is possible to show that the same map can be constructed by directly obtaining the Kraus operators from the Choi matrix corresponding to map $A$ given in Eq. (\ref{eq:Bmat}).
	
	 Let us define the projectors on the system space as $\Pi^{+} = \ket{0_S}\bra{0_S} \otimes \mathbb{1}_E$ and $\Pi^{-} = \ket{1_S}\bra{1_S}  \otimes \mathbb{1}_E$. Here, $\mathbb{1}_E$ is the identity operator on the environment Hilbert space. Applying these projectors on time evolved state of the combined system,  the (normalized) post-measurement states in the two cases are given respectively as:
	
	\begin{align}
	\ket{\phi^o (t)} &= \frac{1}{\sqrt{2}}    \left(
	\begin{array}{c}
	1 \\
	e^{-i \omega t}\\
	0 \\
	0 \\
	\end{array}
	\right) =  \begin{pmatrix}
	1 \\  0
	\end{pmatrix} \otimes \frac{1}{\sqrt{2}} \begin{pmatrix}
	1   \\ e^{-i \omega t}          
	\end{pmatrix}, \\
	\ket{\phi^1 (t)} &=  \frac{1}{\sqrt{2}} \left(
	\begin{array}{c}
	0 \\
	0 \\
	e^{-i \omega t}\\
	1 \\
	\end{array}
	\right) =   \underbrace{\begin{pmatrix}
		0 \\  1
		\end{pmatrix}}_{system} \otimes\underbrace{ \frac{1}{\sqrt{2}} \begin{pmatrix}
		e^{-i \omega t}  \\  1          
		\end{pmatrix}}_{environment}.
	\end{align}   
	
	 Therefore we have two possible evolutions  with the following system and environment states
	\begin{equation} 
	\ket{\chi_S(0) } = \ket{0}, \qquad \ket{\chi_E(0)} = \frac{ \ket{0} + e^{-i \omega t} \ket{1}}{\sqrt{2} }, \qquad~~post~~ \Pi^+~~measurement
	\end{equation}
	and
	\begin{equation}
	\ket{\chi_S(0) } = \ket{1}, \qquad \ket{\chi_E(0)} = \frac{ e^{-i \omega t} \ket{0} +  \ket{1}}{\sqrt{2} },   \qquad~~ post~~ \Pi^-~~measurement.
	\end{equation}   
	
	The corresponding Kraus operators turn out to be
	\begin{align}\label{eq:Kp}
	\mathcal{K}^{+}_{0} (t) = \left(
	\begin{array}{cc}
	\frac{1}{\sqrt{2}} & 0 \\
	-\frac{i e^{-i  \omega t } \sin ( \omega t )}{\sqrt{2}} & \frac{\cos (\omega t  )}{\sqrt{2}} \\
	\end{array}
	\right), \qquad  \mathcal{K}^{+}_{1} (t) = \left(
	\begin{array}{cc}
	\frac{e^{-i \omega t  } \cos ( \omega t )}{\sqrt{2}} & -\frac{i \sin ( \omega t )}{\sqrt{2}} \\
	0 & \frac{e^{-i \omega t }}{\sqrt{2}} \\
	\end{array}
	\right),
	\end{align}      
	and 
	\begin{align}\label{eq:Km}
	\mathcal{K}^{-}_{0} (t) = \left(
	\begin{array}{cc}
	\frac{e^{-i  \omega t}}{\sqrt{2}} & 0 \\
	-\frac{i \sin ( \omega t )}{\sqrt{2}} & \frac{e^{-i  \omega t } \cos ( \omega t )}{\sqrt{2}} \\
	\end{array}
	\right), \qquad  \mathcal{K}^{-}_{1} (t) =  \left(
	\begin{array}{cc}
	\frac{\cos (\omega t )}{\sqrt{2}} & -\frac{i e^{-i \omega t } \sin ( \omega t)}{\sqrt{2}} \\
	0 & \frac{1}{\sqrt{2}} \\
	\end{array}
	\right).
	\end{align}
	
	We denote the corresponding maps by $\Phi^{\pm}$
	\begin{equation}\label{eq:Phipm}
	\Phi^{\pm}[\rho(0)] =\sum\limits_{\mu} \mathcal{K}^{\pm}_{\mu}  \rho(0)( \mathcal{K}_{\mu}^{\pm})^\dagger
	\end{equation}
	Figure (\ref{fig:diag}) summarizes the various steps discussed above.\bigskip

 \subsection{Composition with the  \textit{phasing damping} map}	 
	The Hamiltonian in Eq. (\ref{eq:H}) generates a purely non-Markovian dynamics over the entire time domain as can be concluded from Eq. (\ref{eq:TD}). However, in order study the effect of Markovian versus non-Markovian noise on the degree of violation of LGI, we define a composite map of $\Phi^{\pm}$ and $\Phi^o$ defined  in Eqs. (\ref{eq:Phipm})  and (\ref{eq:Phi}), with the \textit{phase damping}  (PD) noise model. The dynamics is governed by a Markovian map  described by the following Kraus operators: $P_1 = \begin{pmatrix}  1   &  0 \\   0  &  \sqrt{1 - \lambda(t)} \end{pmatrix}$, and $P_2 = \begin{pmatrix}  0   &  0 \\   0  &  \sqrt{ \lambda(t)} \end{pmatrix}$, such that  $\rho =  \begin{pmatrix}  a   &  b \\   c  &  d \end{pmatrix} \rightarrow \rho^\prime = \sum_{i=1,2} P_i \rho P_i^\dagger =  \begin{pmatrix}  a   &  b \sqrt{1 - \lambda(t)} \\   c \sqrt{1 - \lambda(t)}  &  d  \end{pmatrix}$. The  parameterization  $\lambda (t) = 1 - e^{-\Gamma t}$  assures that as $t \rightarrow \infty$, the off-diagonal elements vanish.
	
With  the notation $L_\mu = \{\mathcal{K}_\mu, \mathcal{K}_\mu^{\pm} \}$, we define the composition of the map $\Phi =\{ \Phi^o, \Phi^{\pm}\}$ with the map $\Phi^{PD}$ describing PD dynamics. We denote the composite map $\mathcal{E}_{t \leftarrow t_0} =  [\Phi \circ \Phi^{PD}]_{t \leftarrow t_0}$, such that
\begin{equation}\label{eq:mapE}
\rho(t) = \mathcal{E}_{t \leftarrow t_0} \rho(t_0) = [\Phi \circ \Phi^{PD}]_{t \leftarrow t_0} \rho(t_0)= \sum\limits_{\mu, \nu = 1}^{2} P_\mu L_\nu \rho(t_0) L_\nu^\dagger P_\mu^\dagger. 
\end{equation}
With $\ket{\psi(t_0 = 0)} = \cos(\theta_s) \ket{0} + \sin(\theta_s) \ket{1}$, we have 
\small
\begin{equation}\label{eq:rhot}
\rho(t) = \frac{1}{2} \begin{pmatrix}
                  1 + \cos(\theta_s) \cos^2(\omega t)    &   \sqrt{1- \lambda(t)} \big[ \cos(\omega t) \sin(\theta_s) + i \sin(\omega t) \cos(\theta_s) \big]  \\
                  \sqrt{1- \lambda(t)} \big[ \cos(\omega t) \sin(\theta_s) - i \sin(\omega t) \cos(\theta_s) \big]  &  1 - \cos(\theta_s) \cos^2(\omega t) 
\end{pmatrix}.
\end{equation}
\normalsize

In the limit $\omega \rightarrow 0$ the dynamics reduces  to PD  which is Markovian, while as for $\Gamma \rightarrow 0$ (and hence $\lambda(t) \rightarrow 0$), we obtain  the purely JC-type non-Markovian dynamics. Further, the PD is derived assuming a stationary (infinite) bath, and hence the channel is not modified after intervention.
  \bigskip

\subsection{Classical non-Markovian model} 
Next, we consider a class of semi-Markov processes that exhibit memory in the classical regime and investigate their impact on  LGI violation in the next section.
We take  up a two dimensional system which can jump from one state to another with certain probability. Such a process can be characterized by the following stochastic matrix \cite{classicalNM}
    \begin{equation}
    Q(\tau) = \begin{pmatrix}
                                1- \pi  & \pi \\
                                \pi   &  1- \pi 
    \end{pmatrix} f(\tau) = \Pi f(\tau).
    \end{equation}
  Here, $0 \le \pi \le 1$ is the probability to jump from one site to another and $f(\tau)$ is an arbitrary waiting time distribution and associated survival probability $g(t) = 1 - \int_{0}^{t} d\tau f(\tau)$. It is often convenient to introduce the function $q(t) $ which is an inverse Laplace transform of the quantity $\tilde{q}(u) = \frac{1}{u} \frac{1 - \tilde{f}(u)}{1 + \tilde{f}(u)}$, where $\tilde{x}(u) = \int_{0}^{+\infty} d \tau x(\tau) e^{-u \tau}$, is the Laplace transform of $x(\tau)$.   Such a process can be shown to be Markovian if and only if the waiting time distribution is given by exponential time distributions of the form  $f(\tau) = \lambda e^{-\lambda \tau}$, and non-Markovian otherwise. 

A specific example is now considered in which the waiting time distribution is not an exponential but given as $  f(t) = 2 \frac{p}{s} e^{-st/2} \frac{1}{\xi} \sinh(\xi s t/2) $, with $g(t) = e^{-st/2} \big[ \cosh(\xi st/2) + \frac{1}{\xi}\sinh(\xi st/2)\big]$, and $q(t) = e^{-st/2} \big[ \cosh(\chi st/2) + \frac{1}{\chi} \sinh(\chi st/2)\big]$. This comes from the convolution of two exponential waiting time distributions with different parameters $\lambda_1$ and $\lambda_2$. Here, $s = \lambda_1 + \lambda_2$, $p=\lambda_1 \lambda_2$, $\xi = \sqrt{1 - 4 \frac{p}{s^2}}$, and $\chi = \sqrt{1 - 8 \frac{p}{s^2}}$. A quantum counterpart of the classical semi-Markov process via. a purely dephasing dynamics is governed by the master equation  
                    \begin{equation}
                      \frac{d}{dt} \rho(t)  = \gamma(t) \mathcal{L}_z[\rho], \quad {\rm with}\qquad  \mathcal{L}_z[\rho] = \sigma_z \rho \sigma_z - \rho, \quad {\rm and} \qquad \gamma(t) = -\frac{1}{2} \frac{\dot{q}(t)}{q(t)}.
                    \end{equation}
Such dynamics can be described by a completely positive and trace preserving map characterized by Kraus operators 
\begin{equation}\label{eq:classical}
\mathcal{C}_1 = \sqrt{\frac{1 + q(t)}{2}} \mathbb{1}, \qquad {\rm and}\qquad   \mathcal{C}_2 = \sqrt{\frac{1 - q(t)}{2}} \sigma_z.
\end{equation}
 The map, for $p \leq \frac{s^2}{8}$, as used here, turns out to be Markovian according to the trace distance and divisibility measures despite being non-Markovian classically.

	\begin{figure}[ht!]
		\centering
			\begin{tabular}{cc}
	\includegraphics[width=50mm]{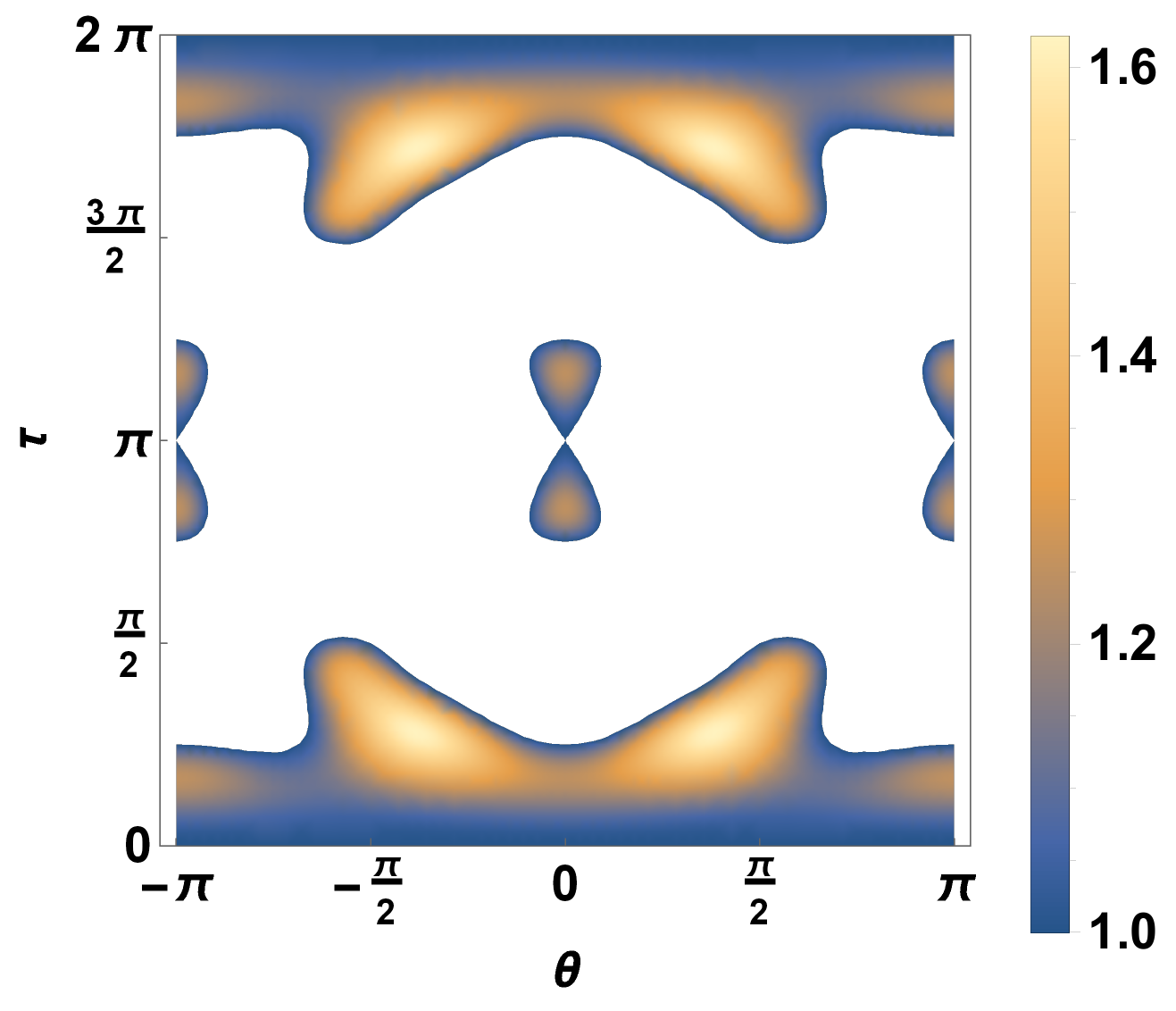} & \includegraphics[width=55mm,height=40mm]{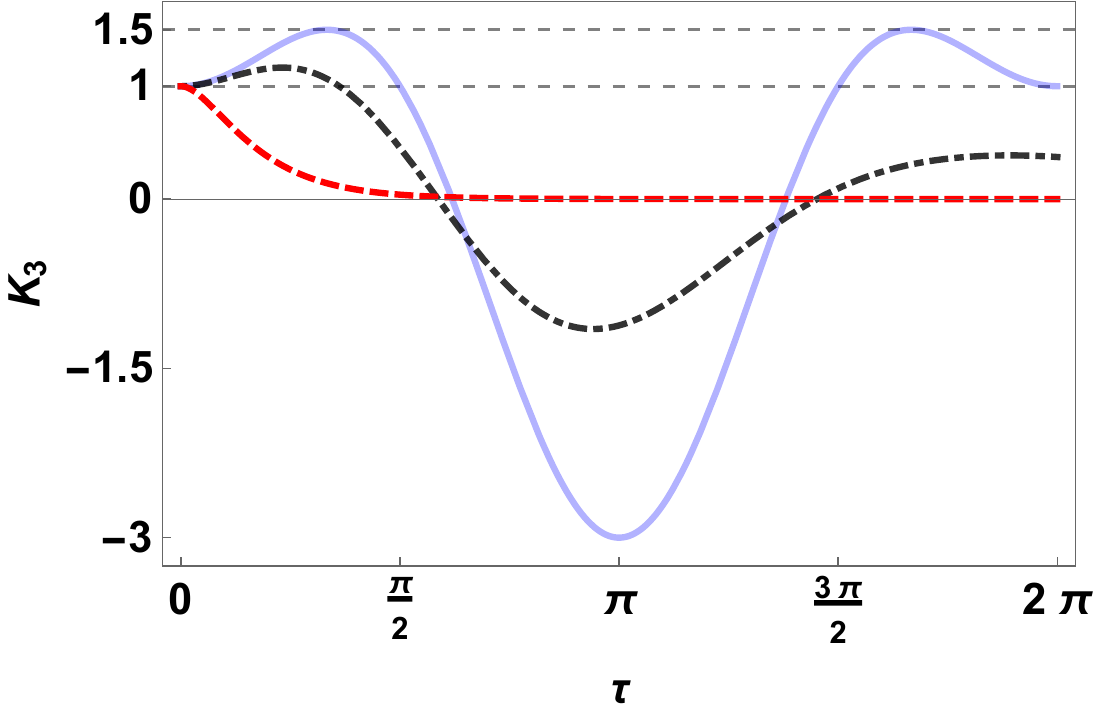}
		\tabularnewline (a)  & (b) \tabularnewline
	 \includegraphics[width=55mm,height=40mm]{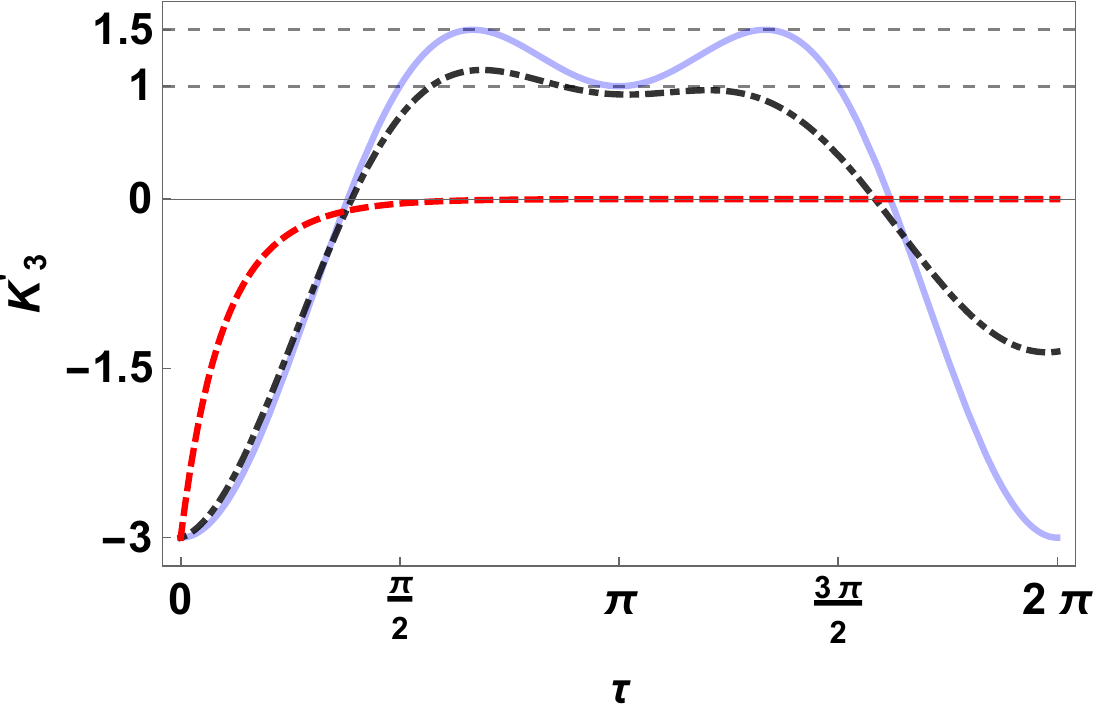} & \includegraphics[width=55mm,height=40mm]{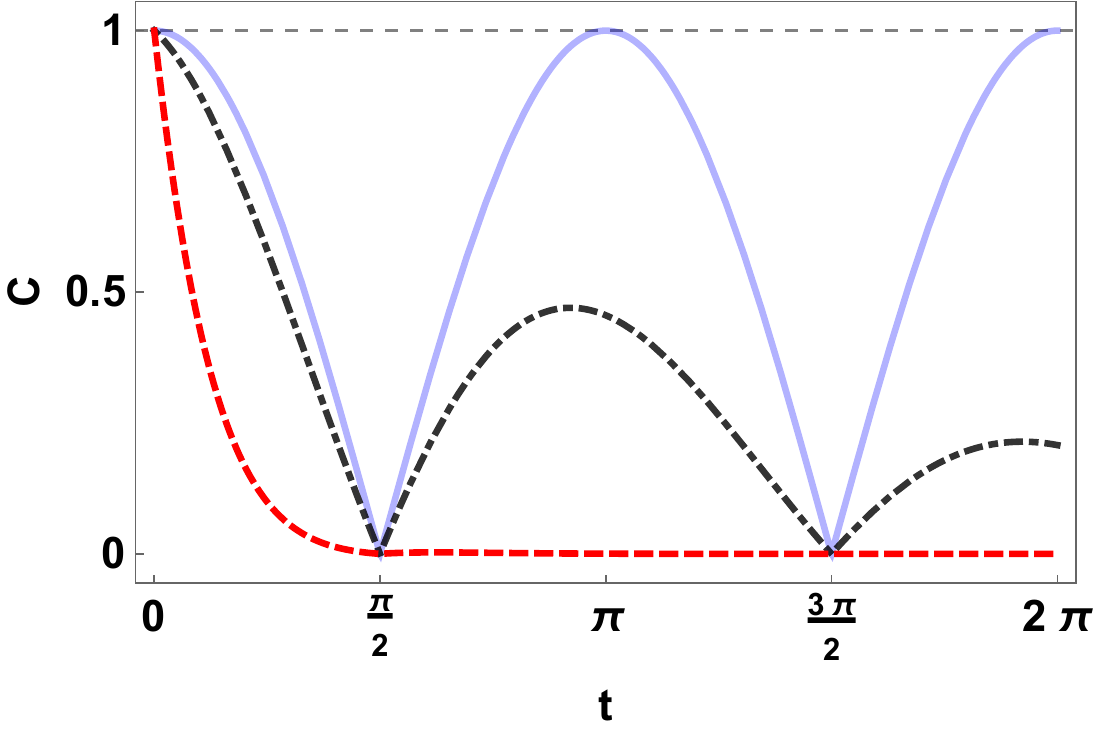} 	
	 	\tabularnewline (c)  & (d) \tabularnewline
	 	\end{tabular}
	\caption{(color online)   (a) The  Leggett-Garg parameter as defined in Eq. (\ref{eq:K3}), for $K_3 > 1$,  with correlation functions given by Eq. (\ref{eq:Cij}), is depicted with respect to time interval $ \uptau$ between the successive measurements and the measurement variable $\theta$ (with $\phi =0$).  The state variables used in this case are $\theta_s = \pi/2,~ \phi_s =0$. Further, we choose $\Gamma = 0$  pertaining to the case when PD  becomes an identity operation and the dynamics is entirely governed  by Hamiltonian $H$ in Eq. (\ref{eq:H}), which generates a non-Markovian subsystem dynamics. The violation reaches to the quantum bound $3/2$ in this case. (b) The parameter $K_3$ is shown for $\Gamma=0$ and $\omega = 1$  with solid (blue) curve. The dot-dashed (black) curve  corresponds to the case when $\omega = 1$ (presence of JC-like model) and  $\Gamma =  0.5$.  The dashed (red) curve shows the scenario when $\omega = 0$ (absence of JC-like model), and $\Gamma = 5$. It is clear that for small $\Gamma$,  the non-Markovian part governed by Hamiltonian $H$ dominates leading to the violation of LGI.  The various parameters used are  $\theta_s = \theta =  \pi/2$, and $ \phi_s = \phi = 0$.  (c) The complementary form of LGI obtained by flipping the sign of observable (as explained in text) leading to $K_3^\prime = - C(0, \uptau) - C(\uptau, 2\uptau) - C(0, 2\uptau) \le 1 $. The various parameters used and the corresponding curve-nomenclature is same as in (b).  (d) The coherence parameter C as defined in Eq. (\ref{eq:coh}), is depicted with respect to  $t$. The characteristic recurrent behavior is observed in non-Markovian regime. The nomenclature of various curves and the parameters used are same as in (b).}
	\label{fig:LGI}
\end{figure}

	\section{Leggett-Garg inequality}\label{LGIs}	
	
 In this section, we study the violation of  the  LGI in the above discussed model. Assuming $t_0 = 0$ and a constant time difference $\uptau$ between successive measurements,  the three time LGI becomes
 	\begin{align}\label{eq:K3}
	K_3 &= C(0, \uptau) + C(\uptau, 2\uptau) - C(0,2\uptau) \le 1,              
	\end{align}
	  where $C(t_i, t_j)$, as defined in Eq. (\ref{Cij}), can be computed as
	 \begin{align}
	  C(t_i, t_j) = \sum_{a,b = \pm} a b~ p(^{a}t_i)q(^{b}t_j|^{a}t_i) &=  \sum_{a,b = \pm} a b \operatorname{Tr}\Big\{\Pi^b \mathcal{E}_{t_j \leftarrow t_i} \big[\Pi^a \rho(t_i) \Pi^{a}\big] \Big\},
	 \end{align}
	  with $\mathcal{E}$ being the composite map defined in Eq. (\ref{eq:mapE}).
 Here, $\Pi^{\pm}$ are the projectors corresponding to a general dichotomic operator 
	 \begin{align}\label{Ok}
	 \operatorname{O}&=
	 \begin{pmatrix}
	 \cos (\theta)            &     e^{i\phi}\sin(\theta)\\
	 e^{-i\phi}\sin(\theta)  &     -\cos (\theta)
	 \end{pmatrix}.                   
	 \end{align}
	 parametrized by $-\pi \le \theta < \pi$; - $\pi/2 \le \phi \le \pi/2$ \cite{murnaghan1962unitary}.  We assume the system is initiated in a pure state $\ket{\psi(0)} = \cos(\theta_s/2) \ket{0} + e^{i\phi_s} \sin(\theta_s/2)\ket{1}$, with $0 \le \theta_s \le \pi$; $0 \le \phi_s < 2 \pi$.  The subsequent dynamics is then governed by the composite map $\mathcal{E}$ defined above. For simplicity, we choose $\phi_s = \phi =0$, and find the  two-time correlation functions 
	\begin{align}\label{eq:Cij}
	C(0, t) &= \cos^2(\theta) \cos^2(\omega \uptau) + \cos(\omega \uptau) \sin^2(\theta) \sqrt{1 - \lambda(\uptau)} \nonumber \\&+ \frac{1}{2} \cos(\theta - \theta_s) \sin(2\theta) \sin^2(\omega \uptau) [\cos(\omega \uptau) - \sqrt{1 - \lambda(\uptau)}] , \nonumber \\
	C(t,2t) &= \cos^2(\omega \uptau) \cos^2(\theta) [1 + \cos(\theta_s) \sin(\theta) \cos(\omega \uptau) \sin^2(\omega \uptau)] \nonumber \\&+ \cos(\omega \uptau) \sin^2(\theta) \sqrt{1 - \lambda(\uptau)} - \frac{1}{2} \cos(\theta + \theta_s) \sin(2\theta) \cos^2(\omega \uptau) \sin^2(\omega \uptau) \sqrt{1 - \lambda (\uptau)} \nonumber \\&+ \frac{1}{2} \sin(\theta) \sin(2\theta) \sin(\theta_s) \cos(\omega \uptau) \sin^2(\omega \uptau) [-1 + \lambda(\uptau)],
	\end{align}
	where $\lambda(\uptau) = 1 - \exp(-\Gamma \uptau)$ as defined in the previous section, is the channel parameter for PD channel. It follows that for $\lambda = 0$ (i.e.,  $\Gamma = 0$), the PD Kraus operators $P_1 =  \mathbb{1}$, $ P_2 = 0$; we call this a trivial operation from the composite map's perspective, i.e., in this case only the evolution generated by Hamiltonian $H$ in Eq. (\ref{eq:H}) is considered. Note that ($\theta_s$, $\phi_s$)  and ($\theta$, $\phi$) are the  state and measurement variables, respectively. For $ \theta = \pi/2$, the  expressions simplify to  $C(0, \uptau) = C( \uptau, 2  \uptau) =  \cos(\omega \uptau) \sqrt{1 - \lambda(\uptau)}$, yielding $K_3 = 2 \cos(\omega \uptau) \sqrt{1 - \lambda(\uptau)} - \cos(2 \omega \uptau) \sqrt{1 - \lambda(2 \uptau)}$, and reaches its maximum quantum bound $3/2$ only if $\lambda(\uptau) = 0$, i.e., when PD is a trivial operation. However, when PD is not a trivial operation, the function $\sqrt{1 - \lambda(\uptau)} = \exp(-\Gamma \uptau/2)$ falls monotonically with time, and therefore reduces the extent of violation such that in the pure Markovian limit, no violation  is observed.  
	
	Figure (\ref{fig:LGI}) depicts the violations of the LGI for various state and measurement settings. Thus for example, from the perspective of LGI violations, we see that given fixed measurement settings, some state preparations are preferable over the others. Also, with fixed state preparation, some measurements are more favorable for the purpose.  Further, the non-Markovian dynamics favors the violation of LGI in comparison to Markovian dynamics, as can be seen, for example, from a comparison of the blue (pure non-Markovian) and the red (pure Markovian) curves in Fig. (\ref{fig:LGI}) (b)-(c).  The enhanced violation of LGI in non-Markovian  regime can be attributed to the information backflow in the non-Markovian case, which counteracts the effect of decoherence, thereby enhancing the quantumness of the system's evolution. This is bolstered by the fact that the coherence shows a recurrent behavior in the non-Markovian case and falls monotonically in the Markovian scenario, as shown in Fig. (\ref{fig:LGI}) (d). The coherence is quantified here by $l_1$-norm 
          \begin{equation}\label{eq:coh}
          C = \sum_{i \ne j}  | \rho_{ij}(t) |,
          \end{equation}    
 where $\rho_{ij} (t)$ is the $ij$-th element of $\rho(t)$ defined in Eq. (\ref{eq:rhot}). The maximum violation of LGI is found to occur at $ \uptau = \pi/3, 5\pi/3$. It is worth mentioning here that a complementary inequality corresponding to Eq. (\ref{eq:K3}) can be obtained by switching the sign of the observable $\operatorname{O} \rightarrow - \operatorname{O}$ leading to $K_3^\prime = - C(0, \uptau) - C(\uptau, 2\uptau) - C(0, 2\uptau) \le 1 $. This inequality shows the maximum violation (not depicted here) at $ \uptau =  2\pi/3, 4\pi/3$. \bigskip
 
\begin{figure}
	\centering
	\includegraphics[width = 70mm]{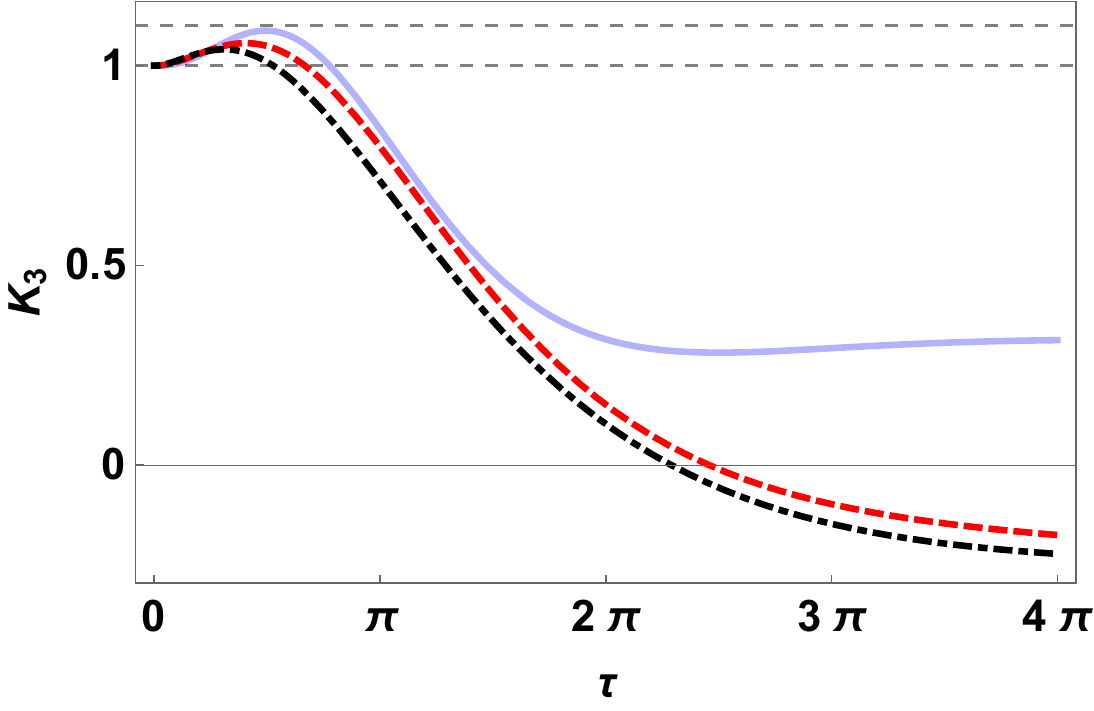}
	\caption{(Color online) Leggett Garg parameter $K_3 = C(0, \uptau) + C(\uptau,  2\uptau) - C(0, 2\uptau)$ with two time correlations given in Eq. (\ref{eq:LGsemiMarkov}), for $\chi = 1$ (blue solid curve), $\chi = 0.5$ (red dashed curve) and $\chi = 0$ (black dot-dashed curve). The parameters used are $\theta_s = \theta = \pi/2$, $\phi_s = \phi = 0$. The maximum violation for these parameters is about $1.1$  compared to $1.5$ observed in case of JC-like model.}
	\label{fig:classical}
\end{figure}

 In order to see the effect of processes that are non-Markovian classically, on the violation of LGI, we make use of the model spelled out in the last section and characterized by the Kraus operators given in Eq. (\ref{eq:classical}). With initial state $\ket{\psi(0)} = \cos(\theta_s) \ket{0} + e^{-i \phi_s} \sin(\theta_s) \ket{1}$ and the general dichotomic observable given in Eq. (\ref{Ok}), the two time correlation functions turn out to be 
 \begin{align}\label{eq:LGsemiMarkov}
     4 C(0, \uptau) &=2 \cos (2 \theta )+\sin (\theta ) \sin (\theta_s) (q(\uptau)+1) \cos (\phi+ \phi_s)  \nonumber \\&+\cos (\theta ) \cos (\theta_s) (q(\uptau)+1)-2 \cos (2 \theta ) q(\uptau)+q(\uptau)+1 , \nonumber \\
  4  C(\uptau, 2\uptau) &= \cos (\theta ) \cos (\theta_s)+2 \cos (2 \theta )+q(\uptau) (\cos (\theta ) \cos (\theta_s) \nonumber \\&-2 \cos (2 \theta )+\sin (\theta ) \sin (\theta_s) (q(\uptau)+1) \cos (\phi + \phi_s)+1)+1.
 \end{align}
 Figure (\ref{fig:classical}) depicts the violations of the LGI for various state and measurement settings, for the quantum semi-Markov process. Violation of LGI is observed in this model, albeit smaller in comparison to the violation of upto maximum quantum bound in a purely quantum non-Markovian system dynamics generated by Hamiltonian $H$ in Eq. (\ref{eq:H}), depicted in Fig. (3) (a).  Although there is no information backflow in this case and thus no recoherence, yet the deviation of the semi-Markovian dynamics from the quantum semigroup structure leads to a relative lowering of decoherence \cite{shrikant2020temporal}, which is conducive to a higher level of violation.

\section{Conclusion}\label{conclusion}
 The violation of the LGI under non-Markovian evolution has been studied by using the reduced dynamics. Difficulties in handling the two-time correlation functions under non-Markovian evolution were highlighted and a possible way of handling them was illustrated by a simple model. The non-Markovian nature of the model was characterized by negative eigenvalues of the Choi matrix implying CP-indivisibility. The  increase in the trace distance function with time brought out the  P-indivisibility of the map.  The non-Markovian dynamics  involves setting up of system-bath correlations; and measurements disrupt these correlations. Therefore, a full system-bath Hamiltonian approach is natural. However, we have pointed out how the problem can be dealt with from a reduced dynamics perspective. The key point is that the noise superoperator acting on the system must be suitably updated after a measurement intervention.   Further, the behavior of LGI violations is compared in Markovain and non-Markovian regimes. It is found that LGI shows violation upto maximum quantum bound in the later case, with no violations in pure Markovian limit. This can be attributed to the fact that non-Markovian dynamics brings memory effects which in turn are exhibited by the recurrent behavior of various quantum features like coherence. We also considered a model in which the underlying classical dynamics is non-Markovian, however, when extended to quantum regime, the dynamical map turns out to be CP divisible and hence Markovian from this perspective. One finds violation of LGI in this model, albeit smaller in comparison to the violation upto maximum quantum bound in a purely quantum non-Markovian model. 


\begin{thebibliography}{63}%
	\makeatletter
	\providecommand \@ifxundefined [1]{%
		\@ifx{#1\undefined}
	}%
	\providecommand \@ifnum [1]{%
		\ifnum #1\expandafter \@firstoftwo
		\else \expandafter \@secondoftwo
		\fi
	}%
	\providecommand \@ifx [1]{%
		\ifx #1\expandafter \@firstoftwo
		\else \expandafter \@secondoftwo
		\fi
	}%
	\providecommand \natexlab [1]{#1}%
	\providecommand \enquote  [1]{``#1''}%
	\providecommand \bibnamefont  [1]{#1}%
	\providecommand \bibfnamefont [1]{#1}%
	\providecommand \citenamefont [1]{#1}%
	\providecommand \href@noop [0]{\@secondoftwo}%
	\providecommand \href [0]{\begingroup \@sanitize@url \@href}%
	\providecommand \@href[1]{\@@startlink{#1}\@@href}%
	\providecommand \@@href[1]{\endgroup#1\@@endlink}%
	\providecommand \@sanitize@url [0]{\catcode `\\12\catcode `\$12\catcode
		`\&12\catcode `\#12\catcode `\^12\catcode `\_12\catcode `\%12\relax}%
	\providecommand \@@startlink[1]{}%
	\providecommand \@@endlink[0]{}%
	\providecommand \url  [0]{\begingroup\@sanitize@url \@url }%
	\providecommand \@url [1]{\endgroup\@href {#1}{\urlprefix }}%
	\providecommand \urlprefix  [0]{URL }%
	\providecommand \Eprint [0]{\href }%
	\providecommand \doibase [0]{http://dx.doi.org/}%
	\providecommand \selectlanguage [0]{\@gobble}%
	\providecommand \bibinfo  [0]{\@secondoftwo}%
	\providecommand \bibfield  [0]{\@secondoftwo}%
	\providecommand \translation [1]{[#1]}%
	\providecommand \BibitemOpen [0]{}%
	\providecommand \bibitemStop [0]{}%
	\providecommand \bibitemNoStop [0]{.\EOS\space}%
	\providecommand \EOS [0]{\spacefactor3000\relax}%
	\providecommand \BibitemShut  [1]{\csname bibitem#1\endcsname}%
	\let\auto@bib@innerbib\@empty
	\bibitem [{\citenamefont {Bell}(1964)}]{bell1964einstein}%
	\BibitemOpen
	\bibfield  {author} {\bibinfo {author} {\bibfnamefont {J.~S.}\ \bibnamefont
			{Bell}},\ }\href {\doibase 10.1103/PhysicsPhysiqueFizika.1.195} {\bibfield
		{journal} {\bibinfo  {journal} {Physics Physique Fizika}\ }\textbf {\bibinfo
			{volume} {1}},\ \bibinfo {pages} {195} (\bibinfo {year} {1964})}\BibitemShut
	{NoStop}%
	\bibitem [{\citenamefont {Einstein}\ \emph {et~al.}(1935)\citenamefont
		{Einstein}, \citenamefont {Podolsky},\ and\ \citenamefont
		{Rosen}}]{PhysRev.47.777}%
	\BibitemOpen
	\bibfield  {author} {\bibinfo {author} {\bibfnamefont {A.}~\bibnamefont
			{Einstein}}, \bibinfo {author} {\bibfnamefont {B.}~\bibnamefont {Podolsky}},
		\ and\ \bibinfo {author} {\bibfnamefont {N.}~\bibnamefont {Rosen}},\ }\href
	{\doibase 10.1103/PhysRev.47.777} {\bibfield  {journal} {\bibinfo  {journal}
			{Phys. Rev.}\ }\textbf {\bibinfo {volume} {47}},\ \bibinfo {pages} {777}
		(\bibinfo {year} {1935})}\BibitemShut {NoStop}%
	\bibitem [{\citenamefont {Schrödinger}(1935)}]{Schordinger1}%
	\BibitemOpen
	\bibfield  {author} {\bibinfo {author} {\bibfnamefont {E.}~\bibnamefont
			{Schrödinger}},\ }\href {\doibase 10.1017/S0305004100013554} {\bibfield
		{journal} {\bibinfo  {journal} {Mathematical Proceedings of the Cambridge
				Philosophical Society}\ }\textbf {\bibinfo {volume} {31}},\ \bibinfo {pages}
		{555–563} (\bibinfo {year} {1935})}\BibitemShut {NoStop}%
	\bibitem [{\citenamefont {Ollivier}\ and\ \citenamefont
		{Zurek}(2001)}]{discordOliverZurek}%
	\BibitemOpen
	\bibfield  {author} {\bibinfo {author} {\bibfnamefont {H.}~\bibnamefont
			{Ollivier}}\ and\ \bibinfo {author} {\bibfnamefont {W.~H.}\ \bibnamefont
			{Zurek}},\ }\href {\doibase 10.1103/PhysRevLett.88.017901} {\bibfield
		{journal} {\bibinfo  {journal} {Phys. Rev. Lett.}\ }\textbf {\bibinfo
			{volume} {88}},\ \bibinfo {pages} {017901} (\bibinfo {year}
		{2001})}\BibitemShut {NoStop}%
	\bibitem [{\citenamefont {Adhikari}\ and\ \citenamefont
		{Banerjee}(2012)}]{DiscordSB}%
	\BibitemOpen
	\bibfield  {author} {\bibinfo {author} {\bibfnamefont {S.}~\bibnamefont
			{Adhikari}}\ and\ \bibinfo {author} {\bibfnamefont {S.}~\bibnamefont
			{Banerjee}},\ }\href {\doibase 10.1103/PhysRevA.86.062313} {\bibfield
		{journal} {\bibinfo  {journal} {Phys. Rev. A}\ }\textbf {\bibinfo {volume}
			{86}},\ \bibinfo {pages} {062313} (\bibinfo {year} {2012})}\BibitemShut
	{NoStop}%
	\bibitem [{\citenamefont {Banerjee}\ \emph {et~al.}(2016)\citenamefont
		{Banerjee}, \citenamefont {Alok},\ and\ \citenamefont
		{MacKenzie}}]{banerjee2016quantum}%
	\BibitemOpen
	\bibfield  {author} {\bibinfo {author} {\bibfnamefont {S.}~\bibnamefont
			{Banerjee}}, \bibinfo {author} {\bibfnamefont {A.~K.}\ \bibnamefont {Alok}},
		\ and\ \bibinfo {author} {\bibfnamefont {R.}~\bibnamefont {MacKenzie}},\
	}\href@noop {} {\bibfield  {journal} {\bibinfo  {journal} {The European
				Physical Journal Plus}\ }\textbf {\bibinfo {volume} {131}},\ \bibinfo {pages}
		{129} (\bibinfo {year} {2016})}\BibitemShut {NoStop}%
	\bibitem [{\citenamefont {Alok}\ \emph {et~al.}(2016)\citenamefont {Alok},
		\citenamefont {Banerjee},\ and\ \citenamefont {Sankar}}]{alok2016quantum}%
	\BibitemOpen
	\bibfield  {author} {\bibinfo {author} {\bibfnamefont {A.~K.}\ \bibnamefont
			{Alok}}, \bibinfo {author} {\bibfnamefont {S.}~\bibnamefont {Banerjee}}, \
		and\ \bibinfo {author} {\bibfnamefont {S.~U.}\ \bibnamefont {Sankar}},\
	}\href@noop {} {\bibfield  {journal} {\bibinfo  {journal} {Nuclear Physics
				B}\ }\textbf {\bibinfo {volume} {909}},\ \bibinfo {pages} {65} (\bibinfo
		{year} {2016})}\BibitemShut {NoStop}%
	\bibitem [{\citenamefont {Banerjee}\ \emph {et~al.}(2015)\citenamefont
		{Banerjee}, \citenamefont {Alok}, \citenamefont {Srikanth},\ and\
		\citenamefont {Hiesmayr}}]{banerjee2015quantum}%
	\BibitemOpen
	\bibfield  {author} {\bibinfo {author} {\bibfnamefont {S.}~\bibnamefont
			{Banerjee}}, \bibinfo {author} {\bibfnamefont {A.~K.}\ \bibnamefont {Alok}},
		\bibinfo {author} {\bibfnamefont {R.}~\bibnamefont {Srikanth}}, \ and\
		\bibinfo {author} {\bibfnamefont {B.~C.}\ \bibnamefont {Hiesmayr}},\
	}\href@noop {} {\bibfield  {journal} {\bibinfo  {journal} {The European
				Physical Journal C}\ }\textbf {\bibinfo {volume} {75}},\ \bibinfo {pages}
		{487} (\bibinfo {year} {2015})}\BibitemShut {NoStop}%
	\bibitem [{\citenamefont {Alok}\ \emph {et~al.}(2015)\citenamefont {Alok},
		\citenamefont {Banerjee},\ and\ \citenamefont {Sankar}}]{Alok:2015iua}%
	\BibitemOpen
	\bibfield  {author} {\bibinfo {author} {\bibfnamefont {A.~K.}\ \bibnamefont
			{Alok}}, \bibinfo {author} {\bibfnamefont {S.}~\bibnamefont {Banerjee}}, \
		and\ \bibinfo {author} {\bibfnamefont {S.~U.}\ \bibnamefont {Sankar}},\
	}\href@noop {} {\bibfield  {journal} {\bibinfo  {journal} {Physics Letters
				B}\ }\textbf {\bibinfo {volume} {749}},\ \bibinfo {pages} {94} (\bibinfo
		{year} {2015})}\BibitemShut {NoStop}%
	\bibitem [{\citenamefont {Chakrabarty}\ \emph {et~al.}(2011)\citenamefont
		{Chakrabarty}, \citenamefont {Banerjee},\ and\ \citenamefont
		{Siddharth}}]{indranilsb}%
	\BibitemOpen
	\bibfield  {author} {\bibinfo {author} {\bibfnamefont {I.}~\bibnamefont
			{Chakrabarty}}, \bibinfo {author} {\bibfnamefont {S.}~\bibnamefont
			{Banerjee}}, \ and\ \bibinfo {author} {\bibfnamefont {N.}~\bibnamefont
			{Siddharth}},\ }\href@noop {} {\bibfield  {journal} {\bibinfo  {journal}
			{Quantum Information and Computation}\ }\textbf {\bibinfo {volume} {11}},\
		\bibinfo {pages} {0541} (\bibinfo {year} {2011})}\BibitemShut {NoStop}%
	\bibitem [{\citenamefont {Banerjee}\ \emph
		{et~al.}(2010{\natexlab{a}})\citenamefont {Banerjee}, \citenamefont
		{Ravishankar},\ and\ \citenamefont {Srikanth}}]{SbRaviSrik}%
	\BibitemOpen
	\bibfield  {author} {\bibinfo {author} {\bibfnamefont {S.}~\bibnamefont
			{Banerjee}}, \bibinfo {author} {\bibfnamefont {V.}~\bibnamefont
			{Ravishankar}}, \ and\ \bibinfo {author} {\bibfnamefont {R.}~\bibnamefont
			{Srikanth}},\ }\href@noop {} {\bibfield  {journal} {\bibinfo  {journal} {The
				European Physical Journal D}\ }\textbf {\bibinfo {volume} {56}},\ \bibinfo
		{pages} {277} (\bibinfo {year} {2010}{\natexlab{a}})}\BibitemShut {NoStop}%
	\bibitem [{\citenamefont {Banerjee}\ \emph
		{et~al.}(2010{\natexlab{b}})\citenamefont {Banerjee}, \citenamefont
		{Ravishankar},\ and\ \citenamefont {Srikanth}}]{SbRaviSrik2}%
	\BibitemOpen
	\bibfield  {author} {\bibinfo {author} {\bibfnamefont {S.}~\bibnamefont
			{Banerjee}}, \bibinfo {author} {\bibfnamefont {V.}~\bibnamefont
			{Ravishankar}}, \ and\ \bibinfo {author} {\bibfnamefont {R.}~\bibnamefont
			{Srikanth}},\ }\href@noop {} {\bibfield  {journal} {\bibinfo  {journal}
			{Annals of Physics}\ }\textbf {\bibinfo {volume} {325}},\ \bibinfo {pages}
		{816} (\bibinfo {year} {2010}{\natexlab{b}})}\BibitemShut {NoStop}%
	\bibitem [{\citenamefont {Dijkstra}\ and\ \citenamefont
		{Tanimura}(2010)}]{Arend}%
	\BibitemOpen
	\bibfield  {author} {\bibinfo {author} {\bibfnamefont {A.~G.}\ \bibnamefont
			{Dijkstra}}\ and\ \bibinfo {author} {\bibfnamefont {Y.}~\bibnamefont
			{Tanimura}},\ }\href@noop {} {\bibfield  {journal} {\bibinfo  {journal}
			{Physical review letters}\ }\textbf {\bibinfo {volume} {104}},\ \bibinfo
		{pages} {250401} (\bibinfo {year} {2010})}\BibitemShut {NoStop}%
	\bibitem [{\citenamefont {Kaer}\ \emph {et~al.}(2010)\citenamefont {Kaer},
		\citenamefont {Nielsen}, \citenamefont {Lodahl}, \citenamefont {Jauho},\ and\
		\citenamefont {M{\o}rk}}]{Kaer}%
	\BibitemOpen
	\bibfield  {author} {\bibinfo {author} {\bibfnamefont {P.}~\bibnamefont
			{Kaer}}, \bibinfo {author} {\bibfnamefont {T.~R.}\ \bibnamefont {Nielsen}},
		\bibinfo {author} {\bibfnamefont {P.}~\bibnamefont {Lodahl}}, \bibinfo
		{author} {\bibfnamefont {A.-P.}\ \bibnamefont {Jauho}}, \ and\ \bibinfo
		{author} {\bibfnamefont {J.}~\bibnamefont {M{\o}rk}},\ }\href@noop {}
	{\bibfield  {journal} {\bibinfo  {journal} {Physical review letters}\
		}\textbf {\bibinfo {volume} {104}},\ \bibinfo {pages} {157401} (\bibinfo
		{year} {2010})}\BibitemShut {NoStop}%
	\bibitem [{\citenamefont {Mirza}(2015)}]{Imran2015}%
	\BibitemOpen
	\bibfield  {author} {\bibinfo {author} {\bibfnamefont {I.~M.}\ \bibnamefont
			{Mirza}},\ }\href@noop {} {\bibfield  {journal} {\bibinfo  {journal} {Journal
				of Modern Optics}\ }\textbf {\bibinfo {volume} {62}},\ \bibinfo {pages}
		{1048} (\bibinfo {year} {2015})}\BibitemShut {NoStop}%
	\bibitem [{\citenamefont {Mirza}\ and\ \citenamefont
		{Schotland}(2016)}]{Imran2016}%
	\BibitemOpen
	\bibfield  {author} {\bibinfo {author} {\bibfnamefont {I.~M.}\ \bibnamefont
			{Mirza}}\ and\ \bibinfo {author} {\bibfnamefont {J.~C.}\ \bibnamefont
			{Schotland}},\ }\href@noop {} {\bibfield  {journal} {\bibinfo  {journal}
			{Physical Review A}\ }\textbf {\bibinfo {volume} {94}},\ \bibinfo {pages}
		{012302} (\bibinfo {year} {2016})}\BibitemShut {NoStop}%
	\bibitem [{\citenamefont {Jiang}\ \emph {et~al.}(2018)\citenamefont {Jiang},
		\citenamefont {Wu},\ and\ \citenamefont {Yang}}]{WeiJiang2018}%
	\BibitemOpen
	\bibfield  {author} {\bibinfo {author} {\bibfnamefont {W.}~\bibnamefont
			{Jiang}}, \bibinfo {author} {\bibfnamefont {F.-Z.}\ \bibnamefont {Wu}}, \
		and\ \bibinfo {author} {\bibfnamefont {G.-J.}\ \bibnamefont {Yang}},\
	}\href@noop {} {\bibfield  {journal} {\bibinfo  {journal} {Physical Review
				A}\ }\textbf {\bibinfo {volume} {98}},\ \bibinfo {pages} {052134} (\bibinfo
		{year} {2018})}\BibitemShut {NoStop}%
	\bibitem [{\citenamefont {Naikoo}\ \emph
		{et~al.}(2018{\natexlab{a}})\citenamefont {Naikoo}, \citenamefont
		{Thapliyal}, \citenamefont {Pathak},\ and\ \citenamefont
		{Banerjee}}]{naikoo2018probing}%
	\BibitemOpen
	\bibfield  {author} {\bibinfo {author} {\bibfnamefont {J.}~\bibnamefont
			{Naikoo}}, \bibinfo {author} {\bibfnamefont {K.}~\bibnamefont {Thapliyal}},
		\bibinfo {author} {\bibfnamefont {A.}~\bibnamefont {Pathak}}, \ and\ \bibinfo
		{author} {\bibfnamefont {S.}~\bibnamefont {Banerjee}},\ }\href@noop {}
	{\bibfield  {journal} {\bibinfo  {journal} {Physical Review A}\ }\textbf
		{\bibinfo {volume} {97}},\ \bibinfo {pages} {063840} (\bibinfo {year}
		{2018}{\natexlab{a}})}\BibitemShut {NoStop}%
	\bibitem [{\citenamefont {Aspect}\ \emph {et~al.}(1981)\citenamefont {Aspect},
		\citenamefont {Grangier},\ and\ \citenamefont
		{Roger}}]{aspect1981experimental}%
	\BibitemOpen
	\bibfield  {author} {\bibinfo {author} {\bibfnamefont {A.}~\bibnamefont
			{Aspect}}, \bibinfo {author} {\bibfnamefont {P.}~\bibnamefont {Grangier}}, \
		and\ \bibinfo {author} {\bibfnamefont {G.}~\bibnamefont {Roger}},\
	}\href@noop {} {\bibfield  {journal} {\bibinfo  {journal} {Physical review
				letters}\ }\textbf {\bibinfo {volume} {47}},\ \bibinfo {pages} {460}
		(\bibinfo {year} {1981})}\BibitemShut {NoStop}%
	\bibitem [{\citenamefont {Tittel}\ \emph {et~al.}(1998)\citenamefont {Tittel},
		\citenamefont {Brendel}, \citenamefont {Gisin}, \citenamefont {Herzog},
		\citenamefont {Zbinden},\ and\ \citenamefont
		{Gisin}}]{tittel1998experimental}%
	\BibitemOpen
	\bibfield  {author} {\bibinfo {author} {\bibfnamefont {W.}~\bibnamefont
			{Tittel}}, \bibinfo {author} {\bibfnamefont {J.}~\bibnamefont {Brendel}},
		\bibinfo {author} {\bibfnamefont {B.}~\bibnamefont {Gisin}}, \bibinfo
		{author} {\bibfnamefont {T.}~\bibnamefont {Herzog}}, \bibinfo {author}
		{\bibfnamefont {H.}~\bibnamefont {Zbinden}}, \ and\ \bibinfo {author}
		{\bibfnamefont {N.}~\bibnamefont {Gisin}},\ }\href@noop {} {\bibfield
		{journal} {\bibinfo  {journal} {Physical Review A}\ }\textbf {\bibinfo
			{volume} {57}},\ \bibinfo {pages} {3229} (\bibinfo {year}
		{1998})}\BibitemShut {NoStop}%
	\bibitem [{\citenamefont {Lanyon}\ \emph {et~al.}(2013)\citenamefont {Lanyon},
		\citenamefont {Jurcevic}, \citenamefont {Hempel}, \citenamefont {Gessner},
		\citenamefont {Vedral}, \citenamefont {Blatt},\ and\ \citenamefont
		{Roos}}]{lanyon2013experimental}%
	\BibitemOpen
	\bibfield  {author} {\bibinfo {author} {\bibfnamefont {B.}~\bibnamefont
			{Lanyon}}, \bibinfo {author} {\bibfnamefont {P.}~\bibnamefont {Jurcevic}},
		\bibinfo {author} {\bibfnamefont {C.}~\bibnamefont {Hempel}}, \bibinfo
		{author} {\bibfnamefont {M.}~\bibnamefont {Gessner}}, \bibinfo {author}
		{\bibfnamefont {V.}~\bibnamefont {Vedral}}, \bibinfo {author} {\bibfnamefont
			{R.}~\bibnamefont {Blatt}}, \ and\ \bibinfo {author} {\bibfnamefont
			{C.}~\bibnamefont {Roos}},\ }\href@noop {} {\bibfield  {journal} {\bibinfo
			{journal} {Physical review letters}\ }\textbf {\bibinfo {volume} {111}},\
		\bibinfo {pages} {100504} (\bibinfo {year} {2013})}\BibitemShut {NoStop}%
	\bibitem [{\citenamefont {Weihs}\ \emph {et~al.}(1998)\citenamefont {Weihs},
		\citenamefont {Jennewein}, \citenamefont {Simon}, \citenamefont
		{Weinfurter},\ and\ \citenamefont {Zeilinger}}]{weihs1998violation}%
	\BibitemOpen
	\bibfield  {author} {\bibinfo {author} {\bibfnamefont {G.}~\bibnamefont
			{Weihs}}, \bibinfo {author} {\bibfnamefont {T.}~\bibnamefont {Jennewein}},
		\bibinfo {author} {\bibfnamefont {C.}~\bibnamefont {Simon}}, \bibinfo
		{author} {\bibfnamefont {H.}~\bibnamefont {Weinfurter}}, \ and\ \bibinfo
		{author} {\bibfnamefont {A.}~\bibnamefont {Zeilinger}},\ }\href@noop {}
	{\bibfield  {journal} {\bibinfo  {journal} {Physical Review Letters}\
		}\textbf {\bibinfo {volume} {81}},\ \bibinfo {pages} {5039} (\bibinfo {year}
		{1998})}\BibitemShut {NoStop}%
	\bibitem [{\citenamefont {Barbieri}(2009)}]{barbieri2009multiple}%
	\BibitemOpen
	\bibfield  {author} {\bibinfo {author} {\bibfnamefont {M.}~\bibnamefont
			{Barbieri}},\ }\href@noop {} {\bibfield  {journal} {\bibinfo  {journal}
			{Physical Review A}\ }\textbf {\bibinfo {volume} {80}},\ \bibinfo {pages}
		{034102} (\bibinfo {year} {2009})}\BibitemShut {NoStop}%
	\bibitem [{\citenamefont {Avis}\ \emph {et~al.}(2010)\citenamefont {Avis},
		\citenamefont {Hayden},\ and\ \citenamefont {Wilde}}]{avis2010leggett}%
	\BibitemOpen
	\bibfield  {author} {\bibinfo {author} {\bibfnamefont {D.}~\bibnamefont
			{Avis}}, \bibinfo {author} {\bibfnamefont {P.}~\bibnamefont {Hayden}}, \ and\
		\bibinfo {author} {\bibfnamefont {M.~M.}\ \bibnamefont {Wilde}},\ }\href@noop
	{} {\bibfield  {journal} {\bibinfo  {journal} {Physical Review A}\ }\textbf
		{\bibinfo {volume} {82}},\ \bibinfo {pages} {030102} (\bibinfo {year}
		{2010})}\BibitemShut {NoStop}%
	\bibitem [{\citenamefont {Lambert}\ \emph {et~al.}(2010)\citenamefont
		{Lambert}, \citenamefont {Emary}, \citenamefont {Chen},\ and\ \citenamefont
		{Nori}}]{lambert2010distinguishing}%
	\BibitemOpen
	\bibfield  {author} {\bibinfo {author} {\bibfnamefont {N.}~\bibnamefont
			{Lambert}}, \bibinfo {author} {\bibfnamefont {C.}~\bibnamefont {Emary}},
		\bibinfo {author} {\bibfnamefont {Y.-N.}\ \bibnamefont {Chen}}, \ and\
		\bibinfo {author} {\bibfnamefont {F.}~\bibnamefont {Nori}},\ }\href@noop {}
	{\bibfield  {journal} {\bibinfo  {journal} {Physical review letters}\
		}\textbf {\bibinfo {volume} {105}},\ \bibinfo {pages} {176801} (\bibinfo
		{year} {2010})}\BibitemShut {NoStop}%
	\bibitem [{\citenamefont {Lambert}\ \emph {et~al.}(2011)\citenamefont
		{Lambert}, \citenamefont {Johansson},\ and\ \citenamefont
		{Nori}}]{lambert2011macrorealism}%
	\BibitemOpen
	\bibfield  {author} {\bibinfo {author} {\bibfnamefont {N.}~\bibnamefont
			{Lambert}}, \bibinfo {author} {\bibfnamefont {R.}~\bibnamefont {Johansson}},
		\ and\ \bibinfo {author} {\bibfnamefont {F.}~\bibnamefont {Nori}},\
	}\href@noop {} {\bibfield  {journal} {\bibinfo  {journal} {Physical Review
				B}\ }\textbf {\bibinfo {volume} {84}},\ \bibinfo {pages} {245421} (\bibinfo
		{year} {2011})}\BibitemShut {NoStop}%
	\bibitem [{\citenamefont {Emary}\ \emph {et~al.}(2013)\citenamefont {Emary},
		\citenamefont {Lambert},\ and\ \citenamefont {Nori}}]{emary2013leggett}%
	\BibitemOpen
	\bibfield  {author} {\bibinfo {author} {\bibfnamefont {C.}~\bibnamefont
			{Emary}}, \bibinfo {author} {\bibfnamefont {N.}~\bibnamefont {Lambert}}, \
		and\ \bibinfo {author} {\bibfnamefont {F.}~\bibnamefont {Nori}},\ }\href@noop
	{} {\bibfield  {journal} {\bibinfo  {journal} {Reports on Progress in
				Physics}\ }\textbf {\bibinfo {volume} {77}},\ \bibinfo {pages} {016001}
		(\bibinfo {year} {2013})}\BibitemShut {NoStop}%
	\bibitem [{\citenamefont {Kofler}\ and\ \citenamefont
		{Brukner}(2013)}]{kofler2013condition}%
	\BibitemOpen
	\bibfield  {author} {\bibinfo {author} {\bibfnamefont {J.}~\bibnamefont
			{Kofler}}\ and\ \bibinfo {author} {\bibfnamefont {{\v{C}}.}~\bibnamefont
			{Brukner}},\ }\href@noop {} {\bibfield  {journal} {\bibinfo  {journal}
			{Physical Review A}\ }\textbf {\bibinfo {volume} {87}},\ \bibinfo {pages}
		{052115} (\bibinfo {year} {2013})}\BibitemShut {NoStop}%
	\bibitem [{\citenamefont {Leggett}\ and\ \citenamefont
		{Garg}(1985)}]{leggett1985quantum}%
	\BibitemOpen
	\bibfield  {author} {\bibinfo {author} {\bibfnamefont {A.~J.}\ \bibnamefont
			{Leggett}}\ and\ \bibinfo {author} {\bibfnamefont {A.}~\bibnamefont {Garg}},\
	}\href@noop {} {\bibfield  {journal} {\bibinfo  {journal} {Physical Review
				Letters}\ }\textbf {\bibinfo {volume} {54}},\ \bibinfo {pages} {857}
		(\bibinfo {year} {1985})}\BibitemShut {NoStop}%
	\bibitem [{\citenamefont {Montina}(2012)}]{montina2012dynamics}%
	\BibitemOpen
	\bibfield  {author} {\bibinfo {author} {\bibfnamefont {A.}~\bibnamefont
			{Montina}},\ }\href@noop {} {\bibfield  {journal} {\bibinfo  {journal}
			{Physical review letters}\ }\textbf {\bibinfo {volume} {108}},\ \bibinfo
		{pages} {160501} (\bibinfo {year} {2012})}\BibitemShut {NoStop}%
	\bibitem [{\citenamefont {Emary}(2012)}]{Emary2012}%
	\BibitemOpen
	\bibfield  {author} {\bibinfo {author} {\bibfnamefont {C.}~\bibnamefont
			{Emary}},\ }\href@noop {} {\bibfield  {journal} {\bibinfo  {journal}
			{Physical Review B}\ }\textbf {\bibinfo {volume} {86}},\ \bibinfo {pages}
		{085418} (\bibinfo {year} {2012})}\BibitemShut {NoStop}%
	\bibitem [{\citenamefont {Emary}(2013)}]{EmaryDecoh}%
	\BibitemOpen
	\bibfield  {author} {\bibinfo {author} {\bibfnamefont {C.}~\bibnamefont
			{Emary}},\ }\href@noop {} {\bibfield  {journal} {\bibinfo  {journal}
			{Physical Review A}\ }\textbf {\bibinfo {volume} {87}},\ \bibinfo {pages}
		{032106} (\bibinfo {year} {2013})}\BibitemShut {NoStop}%
	\bibitem [{\citenamefont {Naikoo}\ \emph {et~al.}(2020)\citenamefont {Naikoo},
		\citenamefont {Alok}, \citenamefont {Banerjee}, \citenamefont {Sankar},
		\citenamefont {Guarnieri}, \citenamefont {Schultze},\ and\ \citenamefont
		{Hiesmayr}}]{Naikoo:2017fos}%
	\BibitemOpen
	\bibfield  {author} {\bibinfo {author} {\bibfnamefont {J.}~\bibnamefont
			{Naikoo}}, \bibinfo {author} {\bibfnamefont {A.~K.}\ \bibnamefont {Alok}},
		\bibinfo {author} {\bibfnamefont {S.}~\bibnamefont {Banerjee}}, \bibinfo
		{author} {\bibfnamefont {S.~U.}\ \bibnamefont {Sankar}}, \bibinfo {author}
		{\bibfnamefont {G.}~\bibnamefont {Guarnieri}}, \bibinfo {author}
		{\bibfnamefont {C.}~\bibnamefont {Schultze}}, \ and\ \bibinfo {author}
		{\bibfnamefont {B.~C.}\ \bibnamefont {Hiesmayr}},\ }\href@noop {} {\bibfield
		{journal} {\bibinfo  {journal} {Nuclear Physics B}\ }\textbf {\bibinfo
			{volume} {951}},\ \bibinfo {pages} {114872} (\bibinfo {year}
		{2020})}\BibitemShut {NoStop}%
	\bibitem [{\citenamefont {Naikoo}\ \emph
		{et~al.}(2018{\natexlab{b}})\citenamefont {Naikoo}, \citenamefont {Alok},\
		and\ \citenamefont {Banerjee}}]{javidLGImeson}%
	\BibitemOpen
	\bibfield  {author} {\bibinfo {author} {\bibfnamefont {J.}~\bibnamefont
			{Naikoo}}, \bibinfo {author} {\bibfnamefont {A.~K.}\ \bibnamefont {Alok}}, \
		and\ \bibinfo {author} {\bibfnamefont {S.}~\bibnamefont {Banerjee}},\
	}\href@noop {} {\bibfield  {journal} {\bibinfo  {journal} {Physical Review
				D}\ }\textbf {\bibinfo {volume} {97}},\ \bibinfo {pages} {053008} (\bibinfo
		{year} {2018}{\natexlab{b}})}\BibitemShut {NoStop}%
	\bibitem [{\citenamefont {Mal}\ \emph {et~al.}(2016)\citenamefont {Mal},
		\citenamefont {Das},\ and\ \citenamefont {Home}}]{mal2016quantum}%
	\BibitemOpen
	\bibfield  {author} {\bibinfo {author} {\bibfnamefont {S.}~\bibnamefont
			{Mal}}, \bibinfo {author} {\bibfnamefont {D.}~\bibnamefont {Das}}, \ and\
		\bibinfo {author} {\bibfnamefont {D.}~\bibnamefont {Home}},\ }\href@noop {}
	{\bibfield  {journal} {\bibinfo  {journal} {Physical Review A}\ }\textbf
		{\bibinfo {volume} {94}},\ \bibinfo {pages} {062117} (\bibinfo {year}
		{2016})}\BibitemShut {NoStop}%
	\bibitem [{\citenamefont {Naikoo}\ and\ \citenamefont
		{Banerjee}(2018)}]{Naikoo2018}%
	\BibitemOpen
	\bibfield  {author} {\bibinfo {author} {\bibfnamefont {J.}~\bibnamefont
			{Naikoo}}\ and\ \bibinfo {author} {\bibfnamefont {S.}~\bibnamefont
			{Banerjee}},\ }\href@noop {} {\bibfield  {journal} {\bibinfo  {journal} {The
				European Physical Journal C}\ }\textbf {\bibinfo {volume} {78}},\ \bibinfo
		{pages} {602} (\bibinfo {year} {2018})}\BibitemShut {NoStop}%
	\bibitem [{\citenamefont {Palacios-Laloy}\ \emph {et~al.}(2010)\citenamefont
		{Palacios-Laloy}, \citenamefont {Mallet}, \citenamefont {Nguyen},
		\citenamefont {Bertet}, \citenamefont {Vion}, \citenamefont {Esteve},\ and\
		\citenamefont {Korotkov}}]{palacios2010experimental}%
	\BibitemOpen
	\bibfield  {author} {\bibinfo {author} {\bibfnamefont {A.}~\bibnamefont
			{Palacios-Laloy}}, \bibinfo {author} {\bibfnamefont {F.}~\bibnamefont
			{Mallet}}, \bibinfo {author} {\bibfnamefont {F.}~\bibnamefont {Nguyen}},
		\bibinfo {author} {\bibfnamefont {P.}~\bibnamefont {Bertet}}, \bibinfo
		{author} {\bibfnamefont {D.}~\bibnamefont {Vion}}, \bibinfo {author}
		{\bibfnamefont {D.}~\bibnamefont {Esteve}}, \ and\ \bibinfo {author}
		{\bibfnamefont {A.~N.}\ \bibnamefont {Korotkov}},\ }\href@noop {} {\bibfield
		{journal} {\bibinfo  {journal} {Nature Physics}\ }\textbf {\bibinfo {volume}
			{6}},\ \bibinfo {pages} {442} (\bibinfo {year} {2010})}\BibitemShut {NoStop}%
	\bibitem [{\citenamefont {Goggin}\ \emph {et~al.}(2011)\citenamefont {Goggin},
		\citenamefont {Almeida}, \citenamefont {Barbieri}, \citenamefont {Lanyon},
		\citenamefont {O’brien}, \citenamefont {White},\ and\ \citenamefont
		{Pryde}}]{goggin2011violation}%
	\BibitemOpen
	\bibfield  {author} {\bibinfo {author} {\bibfnamefont {M.}~\bibnamefont
			{Goggin}}, \bibinfo {author} {\bibfnamefont {M.}~\bibnamefont {Almeida}},
		\bibinfo {author} {\bibfnamefont {M.}~\bibnamefont {Barbieri}}, \bibinfo
		{author} {\bibfnamefont {B.}~\bibnamefont {Lanyon}}, \bibinfo {author}
		{\bibfnamefont {J.}~\bibnamefont {O’brien}}, \bibinfo {author}
		{\bibfnamefont {A.}~\bibnamefont {White}}, \ and\ \bibinfo {author}
		{\bibfnamefont {G.}~\bibnamefont {Pryde}},\ }\href@noop {} {\bibfield
		{journal} {\bibinfo  {journal} {Proceedings of the National Academy of
				Sciences}\ }\textbf {\bibinfo {volume} {108}},\ \bibinfo {pages} {1256}
		(\bibinfo {year} {2011})}\BibitemShut {NoStop}%
	\bibitem [{\citenamefont {Xu}\ \emph {et~al.}(2011)\citenamefont {Xu},
		\citenamefont {Li}, \citenamefont {Zou},\ and\ \citenamefont
		{Guo}}]{xu2011experimental}%
	\BibitemOpen
	\bibfield  {author} {\bibinfo {author} {\bibfnamefont {J.-S.}\ \bibnamefont
			{Xu}}, \bibinfo {author} {\bibfnamefont {C.-F.}\ \bibnamefont {Li}}, \bibinfo
		{author} {\bibfnamefont {X.-B.}\ \bibnamefont {Zou}}, \ and\ \bibinfo
		{author} {\bibfnamefont {G.-C.}\ \bibnamefont {Guo}},\ }\href@noop {}
	{\bibfield  {journal} {\bibinfo  {journal} {Scientific reports}\ }\textbf
		{\bibinfo {volume} {1}},\ \bibinfo {pages} {101} (\bibinfo {year}
		{2011})}\BibitemShut {NoStop}%
	\bibitem [{\citenamefont {Dressel}\ \emph {et~al.}(2011)\citenamefont
		{Dressel}, \citenamefont {Broadbent}, \citenamefont {Howell},\ and\
		\citenamefont {Jordan}}]{dressel2011experimental}%
	\BibitemOpen
	\bibfield  {author} {\bibinfo {author} {\bibfnamefont {J.}~\bibnamefont
			{Dressel}}, \bibinfo {author} {\bibfnamefont {C.}~\bibnamefont {Broadbent}},
		\bibinfo {author} {\bibfnamefont {J.}~\bibnamefont {Howell}}, \ and\ \bibinfo
		{author} {\bibfnamefont {A.~N.}\ \bibnamefont {Jordan}},\ }\href@noop {}
	{\bibfield  {journal} {\bibinfo  {journal} {Physical review letters}\
		}\textbf {\bibinfo {volume} {106}},\ \bibinfo {pages} {040402} (\bibinfo
		{year} {2011})}\BibitemShut {NoStop}%
	\bibitem [{\citenamefont {Suzuki}\ \emph {et~al.}(2012)\citenamefont {Suzuki},
		\citenamefont {Iinuma},\ and\ \citenamefont {Hofmann}}]{suzuki2012violation}%
	\BibitemOpen
	\bibfield  {author} {\bibinfo {author} {\bibfnamefont {Y.}~\bibnamefont
			{Suzuki}}, \bibinfo {author} {\bibfnamefont {M.}~\bibnamefont {Iinuma}}, \
		and\ \bibinfo {author} {\bibfnamefont {H.~F.}\ \bibnamefont {Hofmann}},\
	}\href@noop {} {\bibfield  {journal} {\bibinfo  {journal} {New Journal of
				Physics}\ }\textbf {\bibinfo {volume} {14}},\ \bibinfo {pages} {103022}
		(\bibinfo {year} {2012})}\BibitemShut {NoStop}%
	\bibitem [{\citenamefont {Athalye}\ \emph {et~al.}(2011)\citenamefont
		{Athalye}, \citenamefont {Roy},\ and\ \citenamefont
		{Mahesh}}]{athalye2011investigation}%
	\BibitemOpen
	\bibfield  {author} {\bibinfo {author} {\bibfnamefont {V.}~\bibnamefont
			{Athalye}}, \bibinfo {author} {\bibfnamefont {S.~S.}\ \bibnamefont {Roy}}, \
		and\ \bibinfo {author} {\bibfnamefont {T.}~\bibnamefont {Mahesh}},\
	}\href@noop {} {\bibfield  {journal} {\bibinfo  {journal} {Physical review
				letters}\ }\textbf {\bibinfo {volume} {107}},\ \bibinfo {pages} {130402}
		(\bibinfo {year} {2011})}\BibitemShut {NoStop}%
	\bibitem [{\citenamefont {Katiyar}\ \emph {et~al.}(2013)\citenamefont
		{Katiyar}, \citenamefont {Shukla}, \citenamefont {Rao},\ and\ \citenamefont
		{Mahesh}}]{katiyar2013}%
	\BibitemOpen
	\bibfield  {author} {\bibinfo {author} {\bibfnamefont {H.}~\bibnamefont
			{Katiyar}}, \bibinfo {author} {\bibfnamefont {A.}~\bibnamefont {Shukla}},
		\bibinfo {author} {\bibfnamefont {K.~R.~K.}\ \bibnamefont {Rao}}, \ and\
		\bibinfo {author} {\bibfnamefont {T.}~\bibnamefont {Mahesh}},\ }\href@noop {}
	{\bibfield  {journal} {\bibinfo  {journal} {Physical Review A}\ }\textbf
		{\bibinfo {volume} {87}},\ \bibinfo {pages} {052102} (\bibinfo {year}
		{2013})}\BibitemShut {NoStop}%
	\bibitem [{\citenamefont {Aravinda}\ and\ \citenamefont
		{Srikanth}(2012)}]{aravinda2012general}%
	\BibitemOpen
	\bibfield  {author} {\bibinfo {author} {\bibfnamefont {S.}~\bibnamefont
			{Aravinda}}\ and\ \bibinfo {author} {\bibfnamefont {R.}~\bibnamefont
			{Srikanth}},\ }\href@noop {} {\bibfield  {journal} {\bibinfo  {journal}
			{arXiv:1211.6407}\ } (\bibinfo {year} {2012})}\BibitemShut {NoStop}%
	\bibitem [{\citenamefont {Fritz}(2010)}]{fritz}%
	\BibitemOpen
	\bibfield  {author} {\bibinfo {author} {\bibfnamefont {T.}~\bibnamefont
			{Fritz}},\ }\href@noop {} {\bibfield  {journal} {\bibinfo  {journal} {New
				Journal of Physics}\ }\textbf {\bibinfo {volume} {12}},\ \bibinfo {pages}
		{083055} (\bibinfo {year} {2010})}\BibitemShut {NoStop}%
	\bibitem [{\citenamefont {Kumari}\ and\ \citenamefont
		{Pan}(2017)}]{kumariprob}%
	\BibitemOpen
	\bibfield  {author} {\bibinfo {author} {\bibfnamefont {S.}~\bibnamefont
			{Kumari}}\ and\ \bibinfo {author} {\bibfnamefont {A.}~\bibnamefont {Pan}},\
	}\href@noop {} {\bibfield  {journal} {\bibinfo  {journal} {Physical Review
				A}\ }\textbf {\bibinfo {volume} {96}},\ \bibinfo {pages} {042107} (\bibinfo
		{year} {2017})}\BibitemShut {NoStop}%
	\bibitem [{\citenamefont {Rivas}\ \emph {et~al.}(2014)\citenamefont {Rivas},
		\citenamefont {Huelga},\ and\ \citenamefont {Plenio}}]{rivas2014quantum}%
	\BibitemOpen
	\bibfield  {author} {\bibinfo {author} {\bibfnamefont {A.}~\bibnamefont
			{Rivas}}, \bibinfo {author} {\bibfnamefont {S.~F.}\ \bibnamefont {Huelga}}, \
		and\ \bibinfo {author} {\bibfnamefont {M.~B.}\ \bibnamefont {Plenio}},\
	}\href@noop {} {\bibfield  {journal} {\bibinfo  {journal} {Reports on
				Progress in Physics}\ }\textbf {\bibinfo {volume} {77}},\ \bibinfo {pages}
		{094001} (\bibinfo {year} {2014})}\BibitemShut {NoStop}%
	\bibitem [{\citenamefont {Kumar}\ \emph {et~al.}(2018)\citenamefont {Kumar},
		\citenamefont {Banerjee}, \citenamefont {Srikanth}, \citenamefont
		{Jagadish},\ and\ \citenamefont {Petruccione}}]{pradeep2}%
	\BibitemOpen
	\bibfield  {author} {\bibinfo {author} {\bibfnamefont {N.~P.}\ \bibnamefont
			{Kumar}}, \bibinfo {author} {\bibfnamefont {S.}~\bibnamefont {Banerjee}},
		\bibinfo {author} {\bibfnamefont {R.}~\bibnamefont {Srikanth}}, \bibinfo
		{author} {\bibfnamefont {V.}~\bibnamefont {Jagadish}}, \ and\ \bibinfo
		{author} {\bibfnamefont {F.}~\bibnamefont {Petruccione}},\ }\href@noop {}
	{\bibfield  {journal} {\bibinfo  {journal} {Open Systems \& Information
				Dynamics}\ }\textbf {\bibinfo {volume} {25}},\ \bibinfo {pages} {1850014}
		(\bibinfo {year} {2018})}\BibitemShut {NoStop}%
	\bibitem [{\citenamefont {Omkar}\ \emph {et~al.}(2015)\citenamefont {Omkar},
		\citenamefont {Srikanth},\ and\ \citenamefont
		{Banerjee}}]{omkar2015operator}%
	\BibitemOpen
	\bibfield  {author} {\bibinfo {author} {\bibfnamefont {S.}~\bibnamefont
			{Omkar}}, \bibinfo {author} {\bibfnamefont {R.}~\bibnamefont {Srikanth}}, \
		and\ \bibinfo {author} {\bibfnamefont {S.}~\bibnamefont {Banerjee}},\
	}\href@noop {} {\bibfield  {journal} {\bibinfo  {journal} {Quantum
				Information Processing}\ }\textbf {\bibinfo {volume} {14}},\ \bibinfo {pages}
		{2255} (\bibinfo {year} {2015})}\BibitemShut {NoStop}%
	\bibitem [{\citenamefont {Chen}\ \emph {et~al.}(2016)\citenamefont {Chen},
		\citenamefont {Lambert}, \citenamefont {Li}, \citenamefont {Miranowicz},
		\citenamefont {Chen},\ and\ \citenamefont {Nori}}]{chen2016quantifying}%
	\BibitemOpen
	\bibfield  {author} {\bibinfo {author} {\bibfnamefont {S.-L.}\ \bibnamefont
			{Chen}}, \bibinfo {author} {\bibfnamefont {N.}~\bibnamefont {Lambert}},
		\bibinfo {author} {\bibfnamefont {C.-M.}\ \bibnamefont {Li}}, \bibinfo
		{author} {\bibfnamefont {A.}~\bibnamefont {Miranowicz}}, \bibinfo {author}
		{\bibfnamefont {Y.-N.}\ \bibnamefont {Chen}}, \ and\ \bibinfo {author}
		{\bibfnamefont {F.}~\bibnamefont {Nori}},\ }\href@noop {} {\bibfield
		{journal} {\bibinfo  {journal} {Physical review letters}\ }\textbf {\bibinfo
			{volume} {116}},\ \bibinfo {pages} {020503} (\bibinfo {year}
		{2016})}\BibitemShut {NoStop}%
	\bibitem [{\citenamefont {Goan}\ \emph {et~al.}(2011)\citenamefont {Goan},
		\citenamefont {Chen},\ and\ \citenamefont {Jian}}]{goan2011non}%
	\BibitemOpen
	\bibfield  {author} {\bibinfo {author} {\bibfnamefont {H.-S.}\ \bibnamefont
			{Goan}}, \bibinfo {author} {\bibfnamefont {P.-W.}\ \bibnamefont {Chen}}, \
		and\ \bibinfo {author} {\bibfnamefont {C.-C.}\ \bibnamefont {Jian}},\
	}\href@noop {} {\bibfield  {journal} {\bibinfo  {journal} {The Journal of
				chemical physics}\ }\textbf {\bibinfo {volume} {134}},\ \bibinfo {pages}
		{124112} (\bibinfo {year} {2011})}\BibitemShut {NoStop}%
	\bibitem [{\citenamefont {Chen}\ and\ \citenamefont
		{Ali}(2014)}]{chen2014investigating}%
	\BibitemOpen
	\bibfield  {author} {\bibinfo {author} {\bibfnamefont {P.-W.}\ \bibnamefont
			{Chen}}\ and\ \bibinfo {author} {\bibfnamefont {M.~M.}\ \bibnamefont {Ali}},\
	}\href@noop {} {\bibfield  {journal} {\bibinfo  {journal} {Scientific
				reports}\ }\textbf {\bibinfo {volume} {4}},\ \bibinfo {pages} {6165}
		(\bibinfo {year} {2014})}\BibitemShut {NoStop}%
	\bibitem [{\citenamefont {Ban}(2017)}]{BAN20172313}%
	\BibitemOpen
	\bibfield  {author} {\bibinfo {author} {\bibfnamefont {M.}~\bibnamefont
			{Ban}},\ }\href@noop {} {\bibfield  {journal} {\bibinfo  {journal} {Physics
				Letters A}\ }\textbf {\bibinfo {volume} {381}},\ \bibinfo {pages} {2313}
		(\bibinfo {year} {2017})}\BibitemShut {NoStop}%
	\bibitem [{\citenamefont {Swain}(1981)}]{swain}%
	\BibitemOpen
	\bibfield  {author} {\bibinfo {author} {\bibfnamefont {S.}~\bibnamefont
			{Swain}},\ }\href@noop {} {\bibfield  {journal} {\bibinfo  {journal} {Journal
				of Physics A: Mathematical and General}\ }\textbf {\bibinfo {volume} {14}},\
		\bibinfo {pages} {2577} (\bibinfo {year} {1981})}\BibitemShut {NoStop}%
	\bibitem [{\citenamefont {Guarnieri}\ \emph {et~al.}(2014)\citenamefont
		{Guarnieri}, \citenamefont {Smirne},\ and\ \citenamefont
		{Vacchini}}]{guarnieri2014quantum}%
	\BibitemOpen
	\bibfield  {author} {\bibinfo {author} {\bibfnamefont {G.}~\bibnamefont
			{Guarnieri}}, \bibinfo {author} {\bibfnamefont {A.}~\bibnamefont {Smirne}}, \
		and\ \bibinfo {author} {\bibfnamefont {B.}~\bibnamefont {Vacchini}},\
	}\href@noop {} {\bibfield  {journal} {\bibinfo  {journal} {Physical Review
				A}\ }\textbf {\bibinfo {volume} {90}},\ \bibinfo {pages} {022110} (\bibinfo
		{year} {2014})}\BibitemShut {NoStop}%
	\bibitem [{\citenamefont {Nielsen}\ and\ \citenamefont {Chuang}(2002)}]{NC}%
	\BibitemOpen
	\bibfield  {author} {\bibinfo {author} {\bibfnamefont {M.~A.}\ \bibnamefont
			{Nielsen}}\ and\ \bibinfo {author} {\bibfnamefont {I.}~\bibnamefont
			{Chuang}},\ }\href@noop {} {\enquote {\bibinfo {title} {Quantum computation
				and quantum information},}\ } (\bibinfo {year} {2002})\BibitemShut {NoStop}%
	\bibitem [{\citenamefont {Budini}(2004)}]{budini}%
	\BibitemOpen
	\bibfield  {author} {\bibinfo {author} {\bibfnamefont {A.~A.}\ \bibnamefont
			{Budini}},\ }\href@noop {} {\bibfield  {journal} {\bibinfo  {journal}
			{Physical Review A}\ }\textbf {\bibinfo {volume} {69}},\ \bibinfo {pages}
		{042107} (\bibinfo {year} {2004})}\BibitemShut {NoStop}%
	\bibitem [{\citenamefont {Breuer}\ \emph {et~al.}(2002)\citenamefont {Breuer},
		\citenamefont {Petruccione} \emph {et~al.}}]{breuer}%
	\BibitemOpen
	\bibfield  {author} {\bibinfo {author} {\bibfnamefont {H.-P.}\ \bibnamefont
			{Breuer}}, \bibinfo {author} {\bibfnamefont {F.}~\bibnamefont {Petruccione}},
		\emph {et~al.},\ }\href@noop {} {\emph {\bibinfo {title} {The theory of open
				quantum systems}}}\ (\bibinfo  {publisher} {Oxford University Press on
		Demand},\ \bibinfo {year} {2002})\BibitemShut {NoStop}%
	\bibitem [{\citenamefont {Kraus}\ \emph {et~al.}(1983)\citenamefont {Kraus},
		\citenamefont {B{\"o}hm}, \citenamefont {Dollard},\ and\ \citenamefont
		{Wootters}}]{kraus1983states}%
	\BibitemOpen
	\bibfield  {author} {\bibinfo {author} {\bibfnamefont {K.}~\bibnamefont
			{Kraus}}, \bibinfo {author} {\bibfnamefont {A.}~\bibnamefont {B{\"o}hm}},
		\bibinfo {author} {\bibfnamefont {J.~D.}\ \bibnamefont {Dollard}}, \ and\
		\bibinfo {author} {\bibfnamefont {W.}~\bibnamefont {Wootters}},\ }\href@noop
	{} {\bibfield  {journal} {\bibinfo  {journal} {Lecture notes in physics}\
		}\textbf {\bibinfo {volume} {190}} (\bibinfo {year} {1983})}\BibitemShut
	{NoStop}%
	\bibitem [{\citenamefont {Sudarshan}\ \emph {et~al.}(1961)\citenamefont
		{Sudarshan}, \citenamefont {Mathews},\ and\ \citenamefont {Rau}}]{Sudarshan}%
	\BibitemOpen
	\bibfield  {author} {\bibinfo {author} {\bibfnamefont {E.}~\bibnamefont
			{Sudarshan}}, \bibinfo {author} {\bibfnamefont {P.}~\bibnamefont {Mathews}},
		\ and\ \bibinfo {author} {\bibfnamefont {J.}~\bibnamefont {Rau}},\
	}\href@noop {} {\bibfield  {journal} {\bibinfo  {journal} {Physical Review}\
		}\textbf {\bibinfo {volume} {121}},\ \bibinfo {pages} {920} (\bibinfo {year}
		{1961})}\BibitemShut {NoStop}%
	\bibitem [{\citenamefont {Vacchini}\ \emph {et~al.}(2011)\citenamefont
		{Vacchini}, \citenamefont {Smirne}, \citenamefont {Laine}, \citenamefont
		{Piilo},\ and\ \citenamefont {Breuer}}]{classicalNM}%
	\BibitemOpen
	\bibfield  {author} {\bibinfo {author} {\bibfnamefont {B.}~\bibnamefont
			{Vacchini}}, \bibinfo {author} {\bibfnamefont {A.}~\bibnamefont {Smirne}},
		\bibinfo {author} {\bibfnamefont {E.-M.}\ \bibnamefont {Laine}}, \bibinfo
		{author} {\bibfnamefont {J.}~\bibnamefont {Piilo}}, \ and\ \bibinfo {author}
		{\bibfnamefont {H.-P.}\ \bibnamefont {Breuer}},\ }\href@noop {} {\bibfield
		{journal} {\bibinfo  {journal} {New Journal of Physics}\ }\textbf {\bibinfo
			{volume} {13}},\ \bibinfo {pages} {093004} (\bibinfo {year}
		{2011})}\BibitemShut {NoStop}%
	\bibitem [{\citenamefont {Murnaghan}(1962)}]{murnaghan1962unitary}%
	\BibitemOpen
	\bibfield  {author} {\bibinfo {author} {\bibfnamefont {F.~D.}\ \bibnamefont
			{Murnaghan}},\ }\href@noop {} {\emph {\bibinfo {title} {The unitary and
				rotation groups}}},\ Vol.~\bibinfo {volume} {3}\ (\bibinfo  {publisher}
	{Spartan books},\ \bibinfo {year} {1962})\BibitemShut {NoStop}%
	\bibitem [{\citenamefont {Utagi}\ \emph {et~al.}(2020)\citenamefont {Utagi},
		\citenamefont {Srikanth},\ and\ \citenamefont
		{Banerjee}}]{shrikant2020temporal}%
	\BibitemOpen
	\bibfield  {author} {\bibinfo {author} {\bibfnamefont {S.}~\bibnamefont
			{Utagi}}, \bibinfo {author} {\bibfnamefont {R.}~\bibnamefont {Srikanth}}, \
		and\ \bibinfo {author} {\bibfnamefont {S.}~\bibnamefont {Banerjee}},\
	}\href@noop {} {\bibfield  {journal} {\bibinfo  {journal} {Scientific
				Reports}\ }\textbf {\bibinfo {volume} {10}},\ \bibinfo {pages} {1} (\bibinfo
		{year} {2020})}\BibitemShut {NoStop}%
\end{thebibliography}

%

\end{document}